\documentclass[fleqn,11pt,twoside]{article}
\usepackage{amsthm,amsthm,amssymb, color, epsfig, graphics, subfigure,}
\usepackage{tikz}

\usepackage{amsmath}
\usepackage{graphicx}
\usepackage{amsmath}
\usepackage{graphicx}
\usepackage{latexsym}
\usepackage[all]{xy}
\usepackage{amsthm}
\usepackage{amsmath, amscd}
\usepackage{graphicx}
\usepackage{lscape}
\usepackage{amsmath}
\usepackage{graphicx}
\usepackage{latexsym}

\newtheorem{proposition}{Proposition}

\makeatletter
\newcommand{\copyrightnote}[2]{{\renewcommand{\thefootnote}{}
 \footnotetext{\small\it
\begin{flushleft}
Copyright \copyright \ #1 by  #2
\end{flushleft}}}}

\newcommand{\Name}[1]{\begin{flushleft}
                       \LARGE \bf #1
                       \end{flushleft}\vspace{-3mm}}

\newcommand{\Author}[1]{\begin{flushleft}
                       \it #1 \end{flushleft}}

\newcommand{\Address}[1]{\begin{flushleft}
                       \it #1 \end{flushleft}}

\newcommand{\Date}[1]{\begin{flushleft}
                      \small  \it #1 \end{flushleft}}

\newcommand{\evenhead}{Author \ name}
\newcommand{\oddhead}{Article \ name}

\renewcommand{\@evenhead}{
\hspace*{-3pt}\raisebox{-15pt}[\headheight][0pt]{\vbox{\hbox to \textwidth
{\thepage \hfil \evenhead}\vskip4pt \hrule}}}
\renewcommand{\@oddhead}{
\hspace*{-3pt}\raisebox{-15pt}[\headheight][0pt]{\vbox{\hbox to \textwidth
{\oddhead \hfil \thepage}\vskip4pt\hrule}}}
\renewcommand{\@evenfoot}{}
\renewcommand{\@oddfoot}{}

\long\def\@makecaption#1#2{%
  \vskip\abovecaptionskip
  \sbox\@tempboxa{\small \textbf{#1.}\ \ #2}%
  \ifdim \wd\@tempboxa >\hsize
    {\small \textbf{#1.}\ \ #2}\par
  \else
    \global \@minipagefalse
    \hb@xt@\hsize{\hfil\box\@tempboxa\hfil}%
  \fi
  \vskip\belowcaptionskip}

\newcommand{\JNMPnumberwithin}[3][\arabic]{%
  \@ifundefined{c@#2}{\@nocounterr{#2}}{%
    \@ifundefined{c@#3}{\@nocnterr{#3}}{%
      \@addtoreset{#2}{#3}%
      \@xp\xdef\csname the#2\endcsname{%
        \@xp\@nx\csname the#3\endcsname .\@nx#1{#2}}}}%
}

\newcommand{\resetfootnoterule} {
  \renewcommand\footnoterule{%
  \kern-3\p@
  \hrule\@width.4\columnwidth
  \kern2.6\p@}
}

\renewcommand{\footnoterule}{}

\newcommand{\be}{\begin{equation}}
\newcommand{\ee}{\end{equation}}
\newcommand{\ba}{\hspace*{-5pt}\begin{array}}
\newcommand{\ea}{\end{array}}
\newcommand{\p}{\partial}

\makeatother

\numberwithin{equation}{section}
\theoremstyle{definition}

\renewcommand{\ba}{\begin{array}}
\renewcommand{\ea}{\end{array}}
\newcommand{\beg}{\begin{eqnarray}}
\newcommand{\eeq}{\end{eqnarray}}
\newcommand{\bg}{\begin{eqnarray*}}

\newcommand{\ed}{\end{eqnarray*}}

\newcommand{\nn}{\nonumber}

\renewcommand{\p}{\partial} 
 
\newcommand{\notlhd}{\lhd\kern-.8em{/}\ } 
\newcommand{\notexist}{\ \exists\kern-.5em{\raise.1em\hbox{/}}\ }

\newcommand{\pde}[2]{\frac{\p #1}{\p #2}}

\newcommand{\inp}{{\mbox{\vbox{\hrule width0ex\hbox{\vrule
 height0ex\kern3.8pt
\vbox{\kern2.5pt}\kern3.8pt \vrule height1.6ex}
\hrule width1.6ex}}}}

\begin{document}


\renewcommand{\evenhead}{ {\LARGE\textcolor{blue!10!black!40!green}{{\sf \ \ \ ]ocnmp[}}}\strut\hfill M Euler and N Euler}
\renewcommand{\oddhead}{ {\LARGE\textcolor{blue!10!black!40!green}{{\sf ]ocnmp[}}}\ \ \ \ \   
Potentialisations of some fully-nonlinear symmetry-integrable equations
}


\thispagestyle{empty}
\newcommand{\FistPageHead}[3]{
\begin{flushleft}
\raisebox{8mm}[0pt][0pt]
{\footnotesize \sf
\parbox{150mm}{{Open Communications in Nonlinear Mathematical Physics}\ \  \ {\LARGE\textcolor{blue!10!black!40!green}{]ocnmp[}}
\ \ Vol.4 (2024) pp
#2\hfill {\sc #3}}}\vspace{-13mm}
\end{flushleft}}

\FistPageHead{1}{\pageref{firstpage}--\pageref{lastpage}}{ \ \ Article}

\strut\hfill

\strut\hfill

\copyrightnote{The author(s). Distributed under a Creative Commons Attribution 4.0 International License}

\Name{Potentialisations of a class of fully-nonlinear symmetry-integrable evolution equations}




\label{firstpage}




\Author{Marianna Euler and Norbert Euler $^*$}



\Address{
International Society of Nonlinear Mathematical Physics, 
Auf der Hardt 27,\\
56130 Bad Ems, Germany\\
and\\
Centro Internacional de Ciencias, Av. Universidad s/n, Colonia Chamilpa,\\
 62210 Cuernavaca, Morelos, Mexico\\[0.3cm]
$^*$ Dr.Norbert.Euler@gmail.com
}


\Date{Received March 12, 2024; Accepted May 5, 2024}




\noindent
{\bf Abstract}: 
We consider here the class of fully-nonlinear symmetry-integrable third-order evolution equations in 1+1 dimensions that were proposed recently in {\it Open Communications in Nonlinear Mathematical Physics}, vol. 2, 216--228 (2022). In particular, we report all zero-order and higher-order potentialisations for this class of equations using their integrating factors (or multipliers) up to order four. Chains of connecting evolution equations are also obtained by multi-potentialisations.

\strut\hfill



\renewcommand{\theequation}{\arabic{section}.\arabic{equation}}

\allowdisplaybreaks

\section{Introduction}
We recently reported a class of  fully-nonlinear symmetry-integrable evolution equations in 1+1 dimensions
with rational nonlinearities in their highest derivative \cite{E-E-OCNMP-Dec2022}. This class of equations admit an infinite number of higher-order generalised symmetries, also called Lie-Bäcklund symmetries, and the equations admit recursion operators that generate these symmetries. This class of equations is presented by the following four cases:

\strut\hfill

\noindent
{\proposition
\label{Proposition 1} \cite{E-E-OCNMP-Dec2022}:
The following evolution equations, in the class $u_t=F(u_x,u_{xx},u_{xxx})$ where $F$ a rational function in $u_{xxx}$, are
symmetry-integrable:
\begin{itemize}
\item
{\bf Case I:} The equation
\begin{gather}
\label{Case-I-eq}
u_t=\frac{u_{xx}^6}{(\alpha u_x+\beta)^3u_{xxx}^2}+Q(u_x), 
\end{gather}
where $\{\alpha,\ \beta\}$ are arbitrary constants, not simultaneously zero, and $Q(u_x)$ needs to satisfy
\begin{gather}
\label{Case1-Q-cond}
(\alpha u_x+\beta)\frac{d^5Q}{du_x^5}+5\alpha \frac{d^4Q}{du_x^4}=0,
\end{gather}
which admits for $\alpha\neq 0$ the general solution
\begin{gather}
Q(u_x)=c_5\left(u_x+\frac{\beta}{\alpha}\right)^3+c_4\left(u_x+\frac{\beta}{\alpha}\right)^2
+c_3\left(u_x+\frac{\beta}{\alpha}\right)\nn\\[0.3cm]
\qquad 
+c_2\left(u_x+\frac{\beta}{\alpha}\right)^{-1}
+c_1.
\end{gather}
and for $\alpha=0$, the general solution is
\begin{gather}
Q(u_x)=c_5u_x^4+c_4u_x^3+c_3u_x^2+c_2u_x+c_1.
\end{gather}
Here $c_j$ are arbitrary constants.

\item
{\bf Case II:} The equation
\begin{gather}
\label{Case-II-eq}
u_t=\frac{u_{xx}^3\left(\lambda_1+\lambda_2 u_{xx}\right)^3}{u_{xxx}^2},
\end{gather}
where $\{\lambda_1,\ \lambda_2\}$ are arbitrary constants but not simultaneously zero. 
\item
{\bf Case III:} The equation
\begin{gather}
\label{Case-III-eq}
u_t=\frac{(\alpha u_x+\beta)^{11}}{\left[
\vphantom{\frac{DA}{DB}}
(\alpha u_x+\beta)u_{xxx}-3\alpha u_{xx}^2\right]^2},
\end{gather}
where $\{\alpha,\ \beta\}$ are arbitrary constants but not simultaneously zero.
\item
{\bf Case IV:} The equation
\begin{gather}
\label{Case-IV-eq}
u_t=\frac{4u_x^5}{(2b\,u_x^2-2u_xu_{xxx}+3u_{xx}^2)^2}\equiv \frac{u_x}{(b-S)^2},
\end{gather}
where $b$ is an arbitrary constant and $S$ is the Schwarzian derivative 
\begin{gather}
\label{Schwarzian}
S:=\frac{u_{xxx}}{u_x}-\frac{3}{2}\left(\frac{u_{xx}}{u_x}\right)^2.
\end{gather}

\end{itemize}
}

\strut\hfill

\noindent
The recursion operators for each equation listed in Proposition \ref{Proposition 1} is given in 
\cite{E-E-OCNMP-Dec2022} (regarding recursion operators we refer to \cite{Fokas} and 
\cite{Euler-Euler-book-article}, and the references therein).

The current article is organised as follows: In Section 2 we discuss the method that is applied here for the 
potentialisation of the evolution equations. We state a useful Proposition for the calculation of the conserved currents and define the concept of a higher-order potentialisation, whereby the ``usual'' potentialisation (see for example \cite{Euler-Euler-book-article}) is defined as the {\it zero-order potentialisation} of an evolution equation. In Section 3 we report all potentisalisations of
the equations listed in Proposition 1 and also perform multi-potentialisations where possible. In Section 4 we make some concluding remarks.


\section{Our method using higher-order potentialisations}

In this section we describe the method that is apply to systematically map the equations listed in Proposition 1 by using the equations' conservation laws. The first step in to calculate the equations' integrating factors to obtain, in each case, an appropriate conserved current and flux. We give a short review and introduce our notations.

\strut\hfill

\noindent
{\bf On the notation:} {\it For a given function $\Psi(x,u,u_x,u_{xx},u_{xxx},\ldots,u_{nx})$, where 
$u_{nx}$ denotes the $n$th-derivative of $u$ with respect to $x$, we use the notation $\Psi[x,u]$ to indicate the dependencies.  The {\it order} of $\Psi$ is defined by the highest $x$-derivation in the argument of the function. The same notation is also used to indicate the dependencies of a linear operator.}

\strut\hfill

It is well known that every integrating factor $\Lambda[x,u]$ 
(also known as a multiplier) of a given $n$th-order evolution equation
\begin{gather}
\label{3rd-order-u}
E:=u_t-F[x,u]=0  
\end{gather}
leads to a conserved current $\Phi^t$
and a corresponding flux $\Phi^x$ for (\ref{3rd-order-u}), such that
\begin{gather}
\label{conservation}
\left.\vphantom{\frac{DA}{DB}}
D_t\Phi^t[x,u]+D_x\Phi^x[x,u]\right|_{E=0}=0.
\end{gather}
Now, $\Lambda$ is an integrating factor of (\ref{3rd-order-u}) if and only if
\begin{gather}
\label{E-hat-cond-1}
\hat E[u]\left( \Lambda[x,u]E\right)=0,
\end{gather}
where $\hat E[u]$ denotes the Euler operator
\begin{gather}
\hat E[u]:=\pde{\ }{u}-D_t\circ \pde{\ }{u_t}-D_x\circ \pde{\ }{u_x}
+D_x^2\circ \pde{\ }{u_{xx}}-D_x^3\circ \pde{\ }{u_{xxx}}+\cdots.
\end{gather}
Note that condition (\ref{E-hat-cond-1}) is equivalent to
\begin{subequations}
\begin{gather}
\label{adjoint-symmetry}
\left.\vphantom{\frac{DA}{DB}}
L^*_E[u] \Lambda[x,u]\right|_{E=0}=0\\[0.3cm]
\label{self-adjoint}
\mbox{and}\quad 
L_\Lambda[u]E-L_\Lambda^*[u]E=0.
\end{gather}
\end{subequations}
Here $L_E$ denotes the linear operator
\begin{subequations}
\begin{gather}
L_E[u]=\pde{E}{u}+\pde{E}{u_t}D_t+\pde{E}{u_x}D_x+\pde{E}{u_{xx}}D_x^2+\pde{E}{u_{xxx}}D_x^3+\cdots
\end{gather}
and $L_E^*$ the adjoint of $L_E$, namely
\begin{gather}
L_E^*[u]:=\pde{E}{u}-D_t\circ \pde{E}{u_t}-D_x\circ \pde{E}{u_x}+D_x^2\circ \pde{E}{u_{xx}}
-D_x^3\circ \pde{E}{u_{xxx}}+\cdots.
\end{gather}
\end{subequations}
For more details, we refer to \cite{Anco-Bluman} and \cite{Euler-Euler-book-article}, and the references therein. 
Note that the first condition (\ref{adjoint-symmetry}) requires $\Lambda$ to be an adjoint symmetry for 
(\ref{3rd-order-u}),
whereas the second condition (\ref{self-adjoint}) requires $\Lambda$ to be self-adjoint, which means that
$\Lambda$ must necessarily be even-order. If $\Lambda$ is an adjoint symmetry for (\ref{3rd-order-u}), then, due to the 
linear form of the adjoint symmetry condition (\ref{adjoint-symmetry}), we know that $\Lambda$ must be linear in its highest derivative. This observation is essential (see Proposition 2 below).

The relation between a non-zero integrating factors $\Lambda$ and its corresponding conserved currents $\Phi^t$ for (\ref{3rd-order-u}) is given by the relation
\begin{gather}
\label{Phi-t}
\Lambda[u]=\hat E[u] \Phi^t[x,u],
\end{gather}
whereby the flux $\Phi^x$, say of order $m$, is related to $\Lambda$, $F$ and $\Phi^t$ as follows 
\cite{Euler-Euler-book-article}:
\begin{gather}
\Phi^x[x,u]=-D_x^{-1}(\Lambda F)+\sum_{k=1}^m\sum_{j=0}^{m-k} (-1)^k(D_x^j F)\, D_x^{k-1}
\left(\pde{\Phi^t}{u_{(j+k)}}\right).
\end{gather}

\strut\hfill


Following the relation (\ref{Phi-t}), it is clear that the order of a conserved current $\Phi^t$ for (\ref{3rd-order-u}) is closely related to the order of the corresponding integrating factor $\Lambda$. 
To potentialise an evolution equation (\ref{3rd-order-u}), if at all possible, we make use of the equation's integrating factors whereby the corresponding  lowest order  conserved current $\Phi^t$ are of particular interest. 
In this sense the following statement, which follows directly from (\ref{adjoint-symmetry}) and 
(\ref{self-adjoint}), is useful:

\strut\hfill

\noindent
{\bf Proposition 2:} {\it 
Assume that a given $n$th-order evolution equation of the form (\ref{3rd-order-u}), viz.
\begin{gather*}
u_t=F[x,u],
\end{gather*}
admits an integrating factor $\Lambda$ of order $2m$, where $m$ is a natural number or zero and $n>1$. Then $\Lambda$ is of the form
\begin{gather}
\Lambda[x,u]=g_1(x,u,u_x,\ldots,u_{mx})u_{(2m)x}+g_0(x,u,u_x,\ldots,u_{(2m-1)x})
\end{gather}
and the lowest order 
conserved current $\Phi^t$ of (\ref{3rd-order-u})  is of order $m$, i.e. 
\begin{gather}
\Phi^t[x,u]=\Phi^t(x,u,u_x,\ldots,u_{mx}), 
\end{gather}
whereby
 \begin{gather}
\frac{\p^2\Phi^t}{\ \ \p u_{mx}^2}=(-1)^m g_1(u,u_x,\ldots,u_{mx}).
\end{gather}
}

\strut\hfill

In the current article we are interested in potentialisations of zero and higher-order, which we define as follows: 

\strut\hfill

\noindent
{\bf Definition:} {\it 
Consider an $n$th-order evolution equation of the form (\ref{3rd-order-u}) and assume that it admits an integrating factor $\Lambda$ with corresponding conserved current $\Phi^t$ and flux $\Phi^x$.
\begin{itemize}
\item[a)]
Equation (\ref{3rd-order-u}) is said to be {\bf potentialisable of order zero} if there exist a new dependent variable $v(x,t)$, where
\begin{subequations}
\begin{gather}
v_x:=\Phi^t[x,u],\ \mbox{that is} \\
v_t=-\Phi^x[x,u],
\end{gather}
such that
\begin{gather}
\left. 
\vphantom{\frac{DA}{DB}}
\Phi^x[x,u]\right|_{v_x=\Phi^t}
=G_0(x,v_x,\ldots,v_{nx}).
\end{gather}
The so constructed evolution equation
\begin{gather}
v_t=-G_0(x,v_x,\ldots,v_{nx})
\end{gather}
\end{subequations}
is a {\bf zero-order potential equation} for (\ref{3rd-order-u}), where the explicit form of $G_0$ depends on the given equation
(\ref{3rd-order-u}), on its conserved current $\Phi^t$, and on the corresponding flux $\Phi^x$.

\item[b)]
Equation (\ref{3rd-order-u}) is said to be {\bf potentialisable of order one} if there exist a new dependent variable $w(x,t)$, where
\begin{subequations}
\begin{gather}
w_x:=D_x\Phi^t[x,u],\ \mbox{that is} \\
w_t=-D_x\Phi^x[x,u],
\end{gather}
such that
\begin{gather}
\left. 
\vphantom{\frac{DA}{DB}}
D_x\Phi^x[x,u]\right|_{w_x=D_x\Phi^t}
=G_1(x,w,w_x,\ldots,w_{nx}).
\end{gather}
The so constructed evolution equation
\begin{gather}
w_t=-G_1(x,w,w_x,\ldots,w_{nx})
\end{gather}
\end{subequations}
is a {\bf first-order potential equation} for (\ref{3rd-order-u}),  where the explicit form of $G_1$ depends on the given equation
(\ref{3rd-order-u}), on its conserved current $\Phi^t$, and on the corresponding flux $\Phi^x$.

\item[c)]
Equation (\ref{3rd-order-u}) is said to be {\bf potentialisable of order $p$} if there exist a new dependent variable $w(x,t)$, where $p\geq 1$ and
\begin{subequations}
\begin{gather}
w_x:=D^p_x\Phi^t[x,u],\ \mbox{that is} \\
w_t=-D^p_x\Phi^x[x,u],
\end{gather}
such that
\begin{gather}
\left. 
\vphantom{\frac{DA}{DB}}
D^p_x\Phi^x[x,u]\right|_{w_x=D^p_x\Phi^t}
=G_p(x,w,w_x,\ldots,w_{nx}).
\end{gather}
The so constructed evolution equation
\begin{gather}
w_t=-G_p(x,w,w_x,\ldots,w_{nx})
\end{gather}
\end{subequations}
is a {\bf potential equation of order $p$} for (\ref{3rd-order-u}),  where the explicit form of $G_p$ depends on the given equation
(\ref{3rd-order-u}), on its conserved current $\Phi^t$, and on the corresponding flux $\Phi^x$.

\end{itemize}

}

\noindent
{\bf Remarks:}
\begin{enumerate}
\item 
From the above Definition it is clear that an evolution equation of the form (\ref{3rd-order-u}), which admits a zero-order potentialisation with $v_x=\Phi^t$, will also admit a first-order potentialisation $w_x=D_x\Phi^t$ for the same conserved current $\Phi^t$, whereby $v_x=w$. However, an equation that does not admit a zero-order potentialisation may, or may not, admit a first-order potentialisation. This is also the case for higher-order potentialisations.
\item
Of course, a given equation  (\ref{3rd-order-u})  might not admit a potentialisation of any order for a specific integrating factor. As far as we know, this can not be established {\it a priori} without the knowledge of the integrating factors and the corresponding conservation laws. It is therefore sensible, in our opinion, to do a classification of all integrating factors with all possible corresponding potentialisations that follow, in particular for symmetry-integrable evolution equation. 
\item
In the current work, we are restricting ourselves to integrating factors that do not depend explicitly on their independent variables.
\end{enumerate}

\section{Potentialisations of the equations of Proposition 1}

We now consider each equation listed in Proposition 1, namely Case I, Case II, Case III and Case IV, and derive all possible potentialisations of the fully-nonlinear symmetry-integrable equations listed in this proposition and, where possible, we construct further equations by multi-potentialisations. Diagrams are given in some cases to make the connections between the equations more clear.

\strut\hfill

\noindent
{\bf Case I:} We consider equation (\ref{Case-I-eq}), viz.
\begin{gather*}
u_t=\frac{u_{xx}^6}{(\alpha u_x+\beta)^3u_{xxx}^2}+Q(u_x), 
\end{gather*}
where $\{\alpha,\ \beta\}$ are arbitrary constants, not simultaneously zero, and where $Q(u_x)$ satisfies (\ref{Case1-Q-cond}).

\strut\hfill

\noindent
In its most general form, equation (\ref{Case-I-eq}) admits the following integrating factor of order six:
\begin{gather}
^I\Lambda[u]=\frac{u_{xx}^4u_{6x}}{(\alpha u_x+\beta)^3\,u_{xxx}^3}
+\left(
\frac{12u_{xx}^3}{(\alpha u_x+\beta)^3\,u_{xxx}^2}
-\frac{9\alpha u_{xx}^5}{(\alpha u_x+\beta)^4\,u_{xxx}^3}\right)u_{5x}
\nn\\[0.3cm]
\qquad
+\left(
\frac{24\alpha u_{xx}^2}{(\alpha u_x+\beta)^3\,u_{xxx}}
-\frac{72\alpha u_{xx}^4}{(\alpha u_x+\beta)^4\,u_{xxx}^2}
+\frac{36\alpha^2 u_{xx}^6}{(\alpha u_x+\beta)^5\,u_{xxx}^3}
\right)u_{4x}
\nn\\[0.3cm]
\qquad
-\frac{9u_{xx}^4u_{4x}u_{5x}}{(\alpha u_x+\beta)^3\,u_{xxx}^4}
+\left(
\frac{27\alpha u_{xx}^5}{(\alpha u_x+\beta)^4\,u_{xxx}^4}
-\frac{28u_{xx}^3}{(\alpha u_x+\beta)^3\,u_{xxx}^3}
\right)u_{4x}^2
\nn\\[0.3cm]
\label{Lamb-6th}
\qquad
+\frac{12u_{xx}^4u_{4x}^3}{(\alpha u_x+\beta)^3\,u_{xxx}^5}
-\frac{24u_{xx}u_{xxx}}{(\alpha u_x+\beta)^3}
+\frac{30\alpha^3u_{xx}^7}{(\alpha u_x+\beta)^6\,u_{xxx}^2}
+\frac{108\alpha u_{xx}^3}{(\alpha u_x+\beta)^4}
\nn\\[0.3cm]
\label{Case-I-6O-Lam}
\qquad
-\frac{108\alpha^2u_{xx}^5}{(\alpha u_x+\beta)^5\,u_{xxx}}
-\frac{1}{2}\frac{d^3Q}{du_x^3}\,u_{xx}.
\end{gather}
Here $Q(u_x)$ should satisfy condition (\ref{Case1-Q-cond}), whereby $\alpha$ and $\beta$ are arbitrary constant that are not simultaneously zero.
Using the integrating factor (\ref{Case-I-6O-Lam}) we obtain the following conserved current $\Phi^t$ and flux $\Phi^x$ for  (\ref{Case-I-eq}):
\begin{subequations}
\begin{gather}
\label{Phi-t-6th}
^I\Phi^t[u]=
-\frac{u_{xx}^4}{2(\alpha u_x+\beta)^3u_{xxx}}
+\frac{uu_{xx}}{2u_x^2}\left(\frac{dQ}{du_x} - u_x\frac{d^2Q}{du_x^2}\right)\\[0.3cm]
^I\Phi^x[u]=
\frac{u_{xx}^{10}u_{5x}}{(\alpha u_x+\beta)^6u_{xxx}^5}
-2\frac{u_{xx}^{10}u_{4x}^2}{(\alpha u_x+\beta)^6u_{xxx}^6}
+\frac{uu_{xx}^6u_{4x}}{(\alpha u_x+\beta)^3u_x^2u_{xxx}^3}
\left(\frac{dQ}{du_x}
-u_x\frac{d^2Q}{du_x^2}\right)\nn\\[0.3cm]
\qquad 
+\frac{u_{xx}^9u_{4x}}{(\alpha u_x+\beta)^6u_{xxx}^4}
-\frac{3\alpha u_{xx}^{11}u_{4x}}{(\alpha u_x+\beta)^7u_{xxx}^5}
-\frac{15\alpha^2 u_{xx}^{12}}{4(\alpha u_x+\beta)^8u_{xxx}^4}
+\frac{15\alpha u_{xx}^{10}}{2(\alpha u_x+\beta)^7u_{xxx}^3}\nn\\[0.3cm]
\qquad 
+\frac{1}{2}\left(\frac{dQ}{du_x} - u_x\frac{d^2Q}{du_x^2}\right)
\left(
\frac{3\alpha uu_{xx}^{7}}{(\alpha u_x+\beta)^4u_x^2u_{xxx}^2}
+\frac{u_{xx}^{6}}{(\alpha u_x+\beta)^3u_xu_{xxx}^2}
\right.\nn\\[0.3cm]
\qquad
\left.
-\frac{6uu_{xx}^{5}}{(\alpha u_x+\beta)^3u_x^2u_{xxx}}
\right)
+\frac{1}{2}\frac{dQ}{du_x}\frac{u_{xx}^4}{(\alpha u_x+\beta)^3u_{xxx}}
-\frac{uu_{xx}}{2u_x^2}\frac{dQ}{du_x}\left(\frac{dQ}{du_x} - u_x\frac{d^2Q}{du_x^2}\right)\nn\\[0.3cm]
\label{Phi-x-6th}
\qquad
-\frac{1}{4u_x}\frac{dQ}{du_x}\left(u_x\frac{dQ}{du_x}-2Q\right).
\end{gather}
\end{subequations}
We find that there is no zero-order potentialisation and no higher-order potentialisation related to (\ref{Phi-t-6th}) and 
(\ref{Phi-x-6th}) for any $Q$ that satisfies (\ref{Case1-Q-cond}).

\strut\hfill

We now consider integrating factors of equation (\ref{Case-I-eq}) of order less than six.
Depending on the parameters $\alpha$ and $\beta$ and the form of $Q$, we obtain the following three distinct cases:

\strut\hfill

\noindent
{\bf Subcase I.1:} Let $\alpha\neq 0$ and $\beta\neq 0$. We find that equation (\ref{Case-I-eq})  viz.
\begin{gather*}
u_t=\frac{u_{xx}^6}{(\alpha u_x+\beta)^3u_{xxx}^2}+Q(u_x), 
\end{gather*}
admits  two integrating factors of order four that depend on the form of $Q$, namely for the case $Q^{(2)}=0$ and $Q^{(3)}=0$. There exist no integrating factors of order zero or order two.

\strut\hfill

\noindent
\underline{For $Q^{(2)}=0$}, that is
\begin{gather}
\label{Q-I11-3rd-order}
Q(u_x)=c_1u_x+c_0,
\end{gather}
where $c_0$ and $c_1$ are arbitrary constants of integration, we find that
\begin{gather}
\label{Pot-I11}
u_t=\frac{u_{xx}^6}{(\alpha u_x+\beta)^3u_{xxx}^2}+c_1u_x+c_0
\end{gather}
admits the following two fourth-order integrating factors:
\begin{subequations}
\begin{gather}
\label{Lambda_I1}
^I\Lambda^{1}_{1}[u]=\frac{(\alpha u_x+\beta)u_{4x}}{u_{xx}^3}
-\frac{3(\alpha u_x+\beta)u_{xxx}^2}{u_{xx}^4}
+\frac{2\alpha u_{xxx}}{u_{xx}^2}\\[0.3cm]
\label{Lambda_I2}
^I\Lambda^1_2[u]=\frac{(\alpha u_x+\beta)u_{4x}}{u_{xx}^4}
-\frac{4(\alpha u_x+\beta)u_{xxx}^2}{u_{xx}^5}
+\frac{2\alpha u_{xxx}}{u_{xx}^3}.
\end{gather}
\end{subequations}
Integrating factor (\ref{Lambda_I1}) leads to the
 following conserved current and flux for (\ref{Pot-I11}):
 \begin{subequations}
\begin{gather}
\label{CC-I11}
^I\Phi_{1, 1}^t[u]=\frac{\alpha u_x+\beta}{u_{xx}}
\\[0.3cm]
^I\Phi_{1,1}^x[u]=-\frac{2u_{xx}^4u_{4x}}{(\alpha u_x+\beta)^2u_{xxx}^3}
-\frac{5\alpha u_{xx}^5}{(\alpha u_x+\beta)^3u_{xxx}^2}
+\frac{10u_{xx}^3}{(\alpha u_x+\beta)^2u_{xxx}}\nn\\[0.3cm]
\qquad
-\frac{c_1(\alpha u_x+\beta)}{u_{xx}}
\end{gather}
\end{subequations}
and integrating factor (\ref{Lambda_I2}) gives the 
 following conserved current and flux for (\ref{Pot-I11}):
 \begin{subequations}
 \begin{gather}
 \label{CC-I12}
^I\Phi_{1, 2}^t[u]=\frac{\alpha u_x+\beta}{u_{xx}^2}
\\[0.3cm]
^I\Phi_{1,2}^x[u]=-\frac{4u_{xx}^3u_{4x}}{(\alpha u_x+\beta)^2u_{xxx}^3}
-\frac{9\alpha u_{xx}^4}{(\alpha u_x+\beta)^3u_{xxx}^2}
+\frac{24u_{xx}^2}{(\alpha u_x+\beta)^2u_{xxx}}\nn\\[0.3cm]
\qquad
-\frac{c_1(\alpha u_x+\beta)}{u_{xx}^2}
-\frac{12}{\alpha(\alpha u_x+\beta)}.
\end{gather}
 
\end{subequations}

\noindent
This leads to 

\strut\hfill

\noindent
{\bf Potentialisation I.1:} {\it 
The only conserved current that leads to a potentialisation of  (\ref{Pot-I11}), viz
\begin{gather*}
\boxed{\vphantom{\frac{DA}{DB}}
u_t=\frac{u_{xx}^6}{(\alpha u_x+\beta)^3u_{xxx}^2}+c_1u_x+c_0
}\ ,
\end{gather*}
 is 
$^I\Phi_{1, 1}^t$ given by (\ref{CC-I11}). The zero-order potential equation of (\ref{Pot-I11}) is then
\begin{subequations}
\begin{gather}
v_t=\frac{2v_{xxx}}{(v_{xx}-\alpha)^3}
+\frac{6v_{xx}^2}{(v_{xx}-\alpha)^3v_x}
-\frac{9\alpha v_{xx}}{(v_{xx}-\alpha)^3v_x}
+\frac{3\alpha^2}{(v_{xx}-\alpha)^3v_x}
+c_1 v_x,
\end{gather}
where 
\begin{gather}
v_x=\frac{\alpha u_x+\beta}{u_{xx}}
\end{gather}
and the first-order potential equation of (\ref{Pot-I11}), with $w_x=D_x(^{I}\Phi_{1,1}^t)$, i.e.
$w=v_x$,
is
\begin{gather}
w_t=\frac{2w_{xxx}}{(w_x-\alpha)^3}
-\frac{6w_{xx}^2}{(w_x-\alpha)^4}
-\frac{6w_xw_{xx}}{(w_x-\alpha)^3w}
-\frac{3(2w_x-\alpha)w_x}{8(w_x-\alpha)^2w^2}
+c_1w.
\end{gather}
The conserved current $^I\Phi_{1, 2}^t$ given by  (\ref{CC-I12}) does not give a potentialisation of 
(\ref{Pot-I11}), of any order.

\end{subequations}

}

\strut\hfill

\noindent
\underline{For $Q^{(3)}=0$}, that is
\begin{gather}
\label{Q-I11-3rd-order-N}
Q(u_x)=c_2u_x^2+c_1u_x+c_0,
\end{gather}
where $c_0$, $c_1$ and $c_2\neq 0$ are arbitrary constants with $c_2\neq 0$, we find that
\begin{subequations}
\begin{gather}
\label{Pot-I11_N}
u_t=\frac{u_{xx}^6}{(\alpha u_x+\beta)^3u_{xxx}^2}+c_2u_x^2+c_1u_x+c_0
\end{gather}
admits only one integrating factor of order four, namely (\ref{Lambda_I1}) with the following conserved current and flux:
\begin{gather}
\label{CC-I13}
^I\Phi_{1, 3}^t[u]=\frac{\alpha u_x+\beta}{u_{xx}}
\\[0.3cm]
^I\Phi_{1,3}^x[x,u]=-\frac{2u_{xx}^4u_{4x}}{(\alpha u_x+\beta)^2u_{xxx}^3}
-\frac{5\alpha u_{xx}^5}{(\alpha u_x+\beta)^3u_{xxx}^2}
+\frac{10u_{xx}^3}{(\alpha u_x+\beta)^2u_{xxx}}\nn\\[0.3cm]
\qquad
-\frac{(2c_2u_x+c_1)(\alpha u_x+\beta)}{u_{xx}}
+4c_2(\alpha u+\beta x),
\end{gather}
\end{subequations}

\noindent
We find that equation (\ref{Pot-I11_N}) admits no potentialisation of any order related to the integrating factor (\ref{Lambda_I1}) and its corresponding conserved current (\ref{CC-I13}) for $c_2\neq 0$.

\strut\hfill

\noindent
{\bf Subcase I.2:} Let $\alpha =0$ and $\beta=1$. Equation (\ref{Case-I-eq}) then takes the form
\begin{gather}
\label{SubcaseI2-EQ-Main}
u_t=\frac{u_{xx}^6}{u_{xxx}^2}+c_3u_x^3+c_2u_x^2+c_1u_x+c_0,
\end{gather}
where $c_j$, with $j=0,1,2,3$ are arbitrary constants. We find two distinct cases, namely one case 
where $c_2$ and $c_3$ are both zero, and the case where both $c_2$ are $c_3$ are non-zero.

\strut\hfill

\noindent
\underline{Let $c_2=0$ and $c_3=0$:} The equation (\ref{SubcaseI2-EQ-Main}) then takes the form
\begin{gather}
\label{C-Ia}
u_t=\frac{u_{xx}^6}{u_{xxx}^2}+c_1u_x+c_0.
\end{gather}
Equation  (\ref{C-Ia}) admits the following two fourth-order integrating factors (no zero-order or second-order integrating factors exist):
\begin{subequations}
\begin{gather}
\label{Lambda_I21}
^I\Lambda^{2}_{1}[u]=k_1\left(\frac{u_{4x}}{u_{xx}^3}-\frac{3u_{xxx}^2}{u_{xx}^4}\right)\\[0.3cm]
\label{Lambda_I22}
^I\Lambda^{2}_{2}[u]=k_2\left(\frac{u_{4x}}{u_{xx}^4}-\frac{4u_{xxx}^2}{u_{xx}^5}\right).
\end{gather}
\end{subequations}
Integrating factor (\ref{Lambda_I21}) leads to the
 following conserved current and flux for (\ref{C-Ia}):
 \begin{subequations}
\begin{gather}
\label{CC-I21}
^I\Phi_{2, 1}^t[u]=\frac{k_1}{2u_{xx}}
\\[0.3cm]
^I\Phi_{2,1}^x[u]=k_1\left(-\frac{u_{xx}^4u_{4x}}{u_{xxx}^3}
+\frac{5 u_{xx}^3}{u_{xxx}}
-\frac{c_1}{2u_{xx}}\right).
\end{gather}
\end{subequations}
The second integrating factor (\ref{Lambda_I22}) leads to the 
 following conserved current and flux for (\ref{C-Ia}):
 \begin{subequations}
 \begin{gather}
 \label{CC-I22}
^I\Phi_{2, 2}^t[u]=\frac{k_2}{6u_{xx}^2}
\\[0.3cm]
^I\Phi_{2,2}^x[u]=k_2\left(-\frac{2u_{xx}^3u_{4x}}{3u_{xxx}^3}
+\frac{4 u_{xx}^2}{u_{xxx}}
-\frac{c_1}{6u_{xx}^2}\
+2u_x\right).
\end{gather}
 
 \end{subequations}

\strut\hfill

\noindent
This leads to 

\strut\hfill

\noindent
{\bf Potentialisation I.2:} {\it 

\begin{itemize}
\item[a)]
Using the conserved current  $^I\Phi_{2, 1}^t$, given by (\ref{CC-I21}), we find that
equation (\ref{C-Ia}), viz
\begin{gather*}
\boxed{\vphantom{\frac{DA}{DB}}
u_t=\frac{u_{xx}^6}{u_{xxx}^2}+c_1u_x+c_0
}\ ,
\end{gather*}
admits the zero-order potentiaisation 
\begin{gather}
v_t=\frac{v_{xxx}}{v_{xx}^3}
+\frac{3}{v_xv_{xx}}+c_1v_x,
\end{gather}
where $k_1=2^{2/3}$
\begin{gather}
v_x=\frac{2^{-1/3}}{u_{xx}}
\end{gather}
The first-order potentialisation of (\ref{C-Ia}) with $w_x=D_x(^I\Phi_{2, 1}^t)$ i.e.
$w=v_x$, is then
\begin{gather}
w_t=
\frac{w_{xxx}}{w_x^3}
-3\frac{w_{xx}^2}{w_x^4}
-3\frac{w_{xx}}{ww_x^2}
-\frac{3}{w^2}+c_1w_x.
\end{gather}
\item[b)]
Using the conserved current  $^I\Phi_{2, 2}^t$, given by (\ref{CC-I22}), we find that
equation (\ref{C-Ia}), viz
\begin{gather*}
\boxed{\vphantom{\frac{DA}{DB}}
u_t=\frac{u_{xx}^6}{u_{xxx}^2}+c_1u_x+c_0
}\ ,
\end{gather*}
admits no zero-order potentialisation. However, with
\begin{gather}
w_x=D_x\left(\frac{k_2}{6}\frac{1}{u_{xx}^2}\right)
\end{gather}
we find the following first-order potential equation for (\ref{C-Ia}) in $w$:
\begin{gather}
w_t=\frac{w^{3/2}w_{xxx}}{w_x^3}-\frac{3w^{3/2}w_{xx}^2}{w_x^4}+c_1w_x,
\end{gather}
where $k_2=3\cdot 2^{-5/3}$.
\end{itemize}

}


\strut\hfill

\noindent
\underline{Let $c_3\neq 0$ and $c_2\neq 0$:} The equation (\ref{SubcaseI2-EQ-Main}), viz
\begin{gather*}
u_t=\frac{u_{xx}^6}{u_{xxx}^2}+c_3 u_x^3+c_2u_x^2+c_1u_x+c_0
\end{gather*}
admits one fouth-order integrating factor (none of zero or second-order)
\begin{gather}
\label{Lambda_I31-22}
^I\Lambda^{2}_{3}[u]=\frac{u_{4x}}{u_{xx}^3}-\frac{3u_{xxx}^2}{u_{xx}^4},
\end{gather}
which leads to the following conserved current and flux for  (\ref{SubcaseI2-EQ-Main}):
\begin{subequations}
\begin{gather}
\label{CC-I22b}
^I\Phi_{2, 3}^t[u]=\frac{1}{2u_{xx}^2}
\\[0.3cm]
^I\Phi_{2,3}^x[x,u]=-\frac{u_{xx}^4u_{4x}}{u_{xxx}^3}
+\frac{5 u_{xx}^3}{u_{xxx}}
-\frac{3c_3u_x^2}{2u_{xx}}
-\frac{c_2u_x}{u_{xx}}
-\frac{c_1}{2u_{xx}}
+6c_3u+2c_2x.
\end{gather}
 
 \end{subequations}
 
 \noindent
 Using the conserved current $^I\Phi_{2, 3}^t$ given by (\ref{CC-I22b}),
 we find that equation (\ref{SubcaseI2-EQ-Main}), with $c_3\neq 0$ and $c_2\neq 0$, does not lead to any zero-order or higher-order potentialisation.


\strut\hfill

\noindent
{\bf Subcase I.3:} Let $\alpha=1$ and $\beta=0$. Equation (\ref{Case-I-eq}) then takes the following form:
\begin{gather}
\label{Eq-Case-I.3}
u_t=\frac{u_{xx}^6}{u_x^3u_{xxx}^2}+Q(u_x),\ \mbox{where}\  u_xQ^{(5)}(u_x)+5Q^{(4)}=0.
\end{gather}
We find that (\ref{Eq-Case-I.3}) admits fourth-order integrating factors only in the case where 
\begin{gather}
Q(u_x)=c_2u_x^2+c_1u_x+c_0.
\end{gather}
No zero-order or second-order integrating factors exist for (\ref{Eq-Case-I.3}). Two distinct cases must be considered here, namely the case $c_2=0$ and $c_2\neq 0$.

\strut\hfill

\noindent
\underline{Let $c_2=0$:} Equation (\ref{Eq-Case-I.3}) then takes the form
\begin{gather}
\label{Eq-Case-I.3a}
u_t=\frac{u_{xx}^6}{u_x^3u_{xxx}^2}+c_1u_x+c_0.
\end{gather}
Equation (\ref{Eq-Case-I.3a}) admits the following two fourth-order integrating factors:
\begin{subequations}
\begin{gather}
\label{Lambda_I31}
^I\Lambda^{3}_{1}[u]=2\left(\frac{u_xu_{4x}}{u_{xx}^3}-\frac{3u_xu_{xxx}^2}{u_{xx}^4}+\frac{2u_{xxx}}{u_{xx}^2}\right)\\[0.3cm]
\label{Lambda_I32}
^I\Lambda^{3}_{2}[u]=\frac{u_xu_{4x}}{u_{xx}^4}-\frac{4u_xu_{xxx}^2}{u_{xx}^5}+\frac{2u_{xxx}}{u_{xx}^3}.
\end{gather}
\end{subequations}
Integrating factor (\ref{Lambda_I31}) leads to the following
 conserved current and flux for (\ref{Eq-Case-I.3a}):
 \begin{subequations}
\begin{gather}
\label{CC-I31}
^I\Phi_{3, 1}^t[u]=\frac{u_x}{u_{xx}}
\\[0.3cm]
^I\Phi_{3,1}^x[u]=-\frac{2u_{xx}^4u_{4x}}{u_x^2u_{xxx}^3}
-\frac{5 u_{xx}^5}{u_x^3u_{xxx}^2}
+\frac{10 u_{xx}^3}{u_x^2u_{xxx}}
-\frac{c_1u_x}{u_{xx}}.
\end{gather}
\end{subequations}
and the integrating factor (\ref{Lambda_I32}) gives the 
 following conserved current and flux for (\ref{Eq-Case-I.3a}):
 \begin{subequations}
 \begin{gather}
 \label{CC-I32}
^I\Phi_{3, 2}^t[u]=\frac{u_x}{6u_{xx}^2}
\\[0.3cm]
^I\Phi_{3,2}^x[u]=-\frac{2u_{xx}^3u_{4x}}{3u_x^2u_{xxx}^3}
+\frac{4 u_{xx}^2}{u_x^2u_{xxx}}
-\frac{3u_{xx}^4}{2u_x^3u_{xxx}^2}
-\frac{2}{u_x}
-\frac{c_1u_x}{6u_{xx}^2}.
\end{gather}
 
 \end{subequations}

\noindent
This leads to

\strut\hfill

\noindent
{\bf Potentialisation I.3:} {\it 
Using the conserved current  $^I\Phi_{3, 1}^t$, given by (\ref{CC-I31}), we find that
equation (\ref{Eq-Case-I.3a}), viz
\begin{gather*}
\boxed{\vphantom{\frac{DA}{DB}}
u_t=\frac{u_{xx}^6}{u_x^3u_{xxx}^2}+c_1u_x+c_0
}\ ,
\end{gather*}
admits the zero-order potentiaisation 
\begin{gather}
v_t=\frac{2v_{xxx}}{(v_{xx}-1)^3}
+\frac{3(2v_{xx}-1)}{2v_x(v_{xx}-1)^2}+c_1v_x,
\end{gather}
where
\begin{gather}
v_x=\frac{u_x}{u_{xx}}.
\end{gather}
The first-order potentialisation of (\ref{Eq-Case-I.3a}) with $w_x=D_x(^I\Phi_{3, 1}^t)$ i.e.
$w=v_x$, is then
\begin{gather}
w_t=
\frac{2w_{xxx}}{(w_x-1)^3}
-\frac{6w_{xx}^2}{(w_x-1)^4}
-\frac{6w_xw_{xx}}{w(w_x-1)^3}
-\frac{3w_x(2w_{x}-1)}{w^2(w_x-1)^2}
+c_1w_x.
\end{gather}
No zero or higher-order potentialisations can be obtained for (\ref{Eq-Case-I.3a}) with
the conserved current  $^I\Phi_{3, 2}^t$, given by (\ref{CC-I32}).

}

\strut\hfill

\noindent
\underline{Let $c_2\neq 0$:} The equation
\begin{gather}
\label{Eq-Case-I.3b}
u_t=\frac{u_{xx}^6}{u_x^3u_{xxx}^2}+c_2u_x^2+c_1u_x+c_0
\end{gather}
admits the same fourth-order integrating factor
(\ref{Lambda_I31})
as equation (\ref{Eq-Case-I.3a}) and no zero or second-order integrating factors. This leads to the
 following conserved current and flux for (\ref{Eq-Case-I.3b}):
 \begin{subequations}
\begin{gather}
\label{CC-I33}
^I\Phi_{3, 3}^t[u]=\frac{u_x}{u_{xx}}
\\[0.3cm]
^I\Phi_{3,3}^x[u]=-\frac{2u_{xx}^4u_{4x}}{u_x^2u_{xxx}^3}
-\frac{5 u_{xx}^5}{u_x^3u_{xxx}^2}
+\frac{10 u_{xx}^3}{u_x^2u_{xxx}}
+2c_2\left(\frac{u_{x}^2}{u_{xx}}+2u\right)
-\frac{c_1u_x}{u_{xx}}.
\end{gather}
\end{subequations}
We find that (\ref{Eq-Case-I.3b}) has no zero-order or higher-order potentialisations using (\ref{CC-I33}).

\strut\hfill

\noindent
{\bf Case II:} We consider  (\ref{Case-II-eq}), viz.
\begin{gather*}
u_t=\frac{u_{xx}^3\left(\lambda_1+\lambda_2 u_{xx}\right)^3}{u_{xxx}^2}.
\end{gather*}
Equation (\ref{Case-II-eq}) admits the following four fourth-order integrating factors (no zero-order or second-order 
integrating factors exist for this equation):
\begin{subequations}
\begin{gather}
\label{Case-II-IF-3}
^{II}\Lambda_1[u]=\frac{u_{4x}}{u_{xx}^{3/2}(\lambda_1+\lambda_2u_{xx})^{3/2}}
-\frac{3}{2}\ \frac{(\lambda_1+2\lambda_2 u_{xx})u_{xxx}^2}{u_{xx}^{5/2}(\lambda_1+\lambda_2u_{xx})^{5/2}}\\[0.3cm]
\label{Case-II-IF-1}
^{II}\Lambda_2[u]=\frac{(\lambda_1+2\lambda_2u_{xx})u_{4x}}{u_{xx}^2(\lambda_1+\lambda_2 u_{xx})^2}
-\frac{2\lambda_2(\lambda_1+2\lambda_2u_{xx})u_{xxx}^2}{u_{xx}^2(\lambda_1+\lambda_2 u_{xx})^3}
-\frac{2u_{xxx}^2}{u_{xx}^3(\lambda_1+\lambda_2 u_{xx})}
\\[0.3cm]
\label{Case-II-IF-2}
^{II}\Lambda_3[u]=\frac{u_{4x}}{u_{xx}(\lambda_1+\lambda_2u_{xx})^2}
-\frac{(\lambda_1+3\lambda_2u_{xx})u_{xxx}^2}{u_{xx}^2(\lambda_1+\lambda_2u_{xx})^3}\\[0.3cm]
\label{Case-II-IF-5}
^{II}\Lambda_4[u]=\frac{u_{4x}}{u_{xx}^2(\lambda_1+\lambda_2u_{xx})}
-\frac{(2\lambda_1+3\lambda_2u_{xx})u_{xxx}^2}{u_{xx}^3(\lambda_1+\lambda_2 u_{xx})^2}.
\end{gather}
\end{subequations}

\noindent
The four listed integrating factors, (\ref{Case-II-IF-3}) - (\ref{Case-II-IF-5}), lead four essentially different subcases, which we now discuss in detail under Subcase II.1 to Subcase II.4 below.

\strut\hfill

\noindent
{\bf Subcase II.1:} We consider the integrating factor $^{II}\Lambda_1$ given by (\ref{Case-II-IF-3}) for equation (\ref{Case-II-eq}), viz.
\begin{gather*}
u_t=\frac{u_{xx}^3\left(\lambda_1+\lambda_2 u_{xx}\right)^3}{u_{xxx}^2}.
\end{gather*}
We now discuss the case $\lambda_1\neq 0$ and $\lambda_2\neq 0$, the case $\lambda_1= 0$ and 
$\lambda_2=1$, and the case $\lambda_1= 1$ and 
$\lambda_2=0$ below.

\strut\hfill

\noindent
\underline{Let $\lambda_1\neq 0$ and $\lambda_2\neq 0$:} 
The integrating factor (\ref{Case-II-IF-3}) then leads to the following conserved current and flux for 
(\ref{Case-II-eq}):
\begin{subequations}
\begin{gather}
\label{Phi-t-II-lam-1-1}
^{II}\Phi^t_{1,1}[u]=-\frac{2}{\lambda_1^2}u_{xx}^{1/2}(\lambda_1+\lambda_2 u_{xx})^{1/2}\\[0.3cm]
^{II}\Phi^x_{1,1}[u]=
-\frac{2}{\lambda_1^2}(\lambda_1+2\lambda_2 u_{xx})(\lambda_1+\lambda_2 u_{xx})^{5/2}
\frac{u_{4x}}{u_{xxx}^3}\nn\\
\qquad
+\frac{4}{\lambda_1^2}(\lambda_1+\lambda_2 u_{xx})^{3/2}(\lambda_1^2
+3\lambda_1\lambda_2 u_{xx}+3\lambda_2^2u_{xx}^2)\frac{u_{xx}^{3/2}}{u_{xxx}}.
\end{gather}
\end{subequations}

\smallskip

\noindent
\underline{Let $\lambda_1=0$ and $\lambda_2=1$:} 
Equation (\ref{Case-II-eq}) then takes the form
\begin{gather}
\label{Case-II-eq-lam-0-1-N}
u_t=\frac{u_{xx}^6}{u_{xxx}^2}
\end{gather}
and the integrating factor (\ref{Case-II-IF-3}) leads to the following conserved current and flux for 
(\ref{Case-II-eq-lam-0-1-N}):
\begin{subequations}
\begin{gather}
\label{Phi-t-II-lam-0-1}
^{II}\Phi^t_{1,2}[u]=\frac{2^{-1/3}}{u_{xx}}\\[0.3cm]
\label{Phi-x-II-lam-0-1}
^{II}\Phi^x_{1,2}[u]=
2^{2/3}\left(-\frac{u_{xx}^4u_{4x}}{u_{xxx}^3}+\frac{5u_{xx}^3}{u_{xxx}}\right).
\end{gather}
\end{subequations}

\smallskip

\noindent
\underline{Let $\lambda_1\neq 0$ and $\lambda_2=0$:} 
Equation (\ref{Case-II-eq}) then takes the form
\begin{gather}
\label{Case-II-eq-lam-1-0-N}
u_t=\lambda_1^3\frac{u_{xx}^3}{u_{xxx}^2}
\end{gather}
and the integrating factor (\ref{Case-II-IF-3}) leads to the following conserved current and flux for 
(\ref{Case-II-eq-lam-1-0-N}):
\begin{subequations}
\begin{gather}
\label{Phi-t-II-lam-1-0}
^{II}\Phi^t_{1,3}[u]=\lambda_1^{-3/2}u_{xx}^{1/2}\\[0.3cm]
\label{Phi-x-II-lam-1-0}
^{II}\Phi^x_{1,3}[u]=\lambda^{3/2}\left(\frac{u_{xx}^{5/2}u_{4x}}{u_{xxx}^3}-\frac{2u_{xx}^{3/2}}{u_{xxx}}\right).
\end{gather}
\end{subequations}

\noindent
This leads to 

\strut\hfill

\noindent
{\bf Potentialisation II.1} {\it 
\begin{itemize}
\item[a)] Using the conserved current $^{II}\Phi^t_{1,1}$,
given by (\ref{Phi-t-II-lam-1-1}), we find that equation (\ref{Case-II-eq}) viz.
\begin{gather*}
\boxed{\vphantom{\frac{DA}{DB}}
u_t=\frac{u_{xx}^3\left(\lambda_1+\lambda_2 u_{xx}\right)^3}{u_{xxx}^2}
}
\end{gather*}
admits the zero-order potentialisation
\begin{gather}
v_t=\frac{\lambda_1^3}{4}
\left(
\vphantom{\frac{DA}{DB}}
1+\lambda_1^2\lambda_2v_x^2\right)^{3/2}
\left(\frac{v_x^3v_{xxx}}{v_{xx}^3}-\frac{3v_x^2}{v_{xx}}\right),
\end{gather}
where 
\begin{gather*}
v_x=
-\frac{2}{\lambda_1^2}\left[u_{xx}(\lambda_1+\lambda_2 u_{xx})\right]^{1/2}.
\end{gather*}
The first-order potentialisation for (\ref{Case-II-eq})  with $w_x=D_x(^{II}\Phi_{1, 1}^t)$ i.e.
$w=v_x$, is then
\begin{gather}
w_t=\left(\frac{\lambda_1^3}{4}\right)
\frac{(1+\lambda_1^2\lambda_2 w^2)^{3/2}w^3w_{xxx}}{w_x^3}
-\left(\frac{3\lambda_1^3}{4}\right)
\frac{(1+\lambda_1^2\lambda_2 w^2)^{3/2}w^3w_{xx}^2}{w_x^4}\nn\\[0.3cm]
\qquad
+\left(\frac{3\lambda_1^3}{4}\right)
\frac{ (1+\lambda_1^2\lambda_2 w^2)^{1/2}(2+3\lambda_1^2\lambda_2w^2) w^2w_{xx}}{w_x^2}\nn\\[0.3cm]
\qquad
-\left(\frac{3\lambda_1^3}{4}\right)
(1+\lambda_1^2\lambda_2 w^2)^{1/2}(2+5\lambda_1^2\lambda_2w^2)w.
\end{gather}

\item[b)] Using the conserved current $^{II}\Phi^t_{1,2}$,
given by (\ref{Phi-t-II-lam-0-1}), we find that equation (\ref{Case-II-eq-lam-0-1-N})
viz.
\begin{gather*}
\boxed{\vphantom{\frac{DA}{DB}}
u_t=\frac{u_{xx}^6}{u_{xxx}^2}
}
\end{gather*}
admits the zero-order potentialisation
\begin{gather}
\label{Case-II-b-v}
v_t=\frac{v_{xxx}}{v_{xx}^3}+\frac{3}{v_xv_{xx}},
\end{gather}
where 
\begin{gather*}
v_x=\frac{1}{2^{1/3}u_{xx}}.
\end{gather*}
The first-order potentialisation for (\ref{Case-II-eq-lam-0-1-N})  with $w_x=D_x(^{II}\Phi_{1, 2}^t)$ i.e.
$w=v_x$, is then
\begin{gather}
\label{Case-II-b-w}
w_t=\frac{w_{xxx}}{w_x^3}-\frac{3w_{xx}^2}{w_x^4}-\frac{3w_{xx}}{w w_x^2}-\frac{3}{w^2}.
\end{gather}

\noindent
Turning to the multi-potentialisation of  (\ref{Case-II-eq-lam-0-1-N}), we find the zero-order potentialisation of
 (\ref{Case-II-b-v}) in
\begin{gather}
\label{Case-II-b-v-vt}
\tilde v_t=\frac{\tilde v_x^6\tilde v_{xxx}}{\tilde v_{xx}^3}-3\frac{\tilde v_x^5}{\tilde v_{xx}},
\end{gather}
where 
\begin{gather*}
\tilde v_x=-\frac{1}{v_x},
\end{gather*}
as well as the zero-order potentialisation of (\ref{Case-II-b-v-vt}) in
\begin{gather}
\label{Case-II-b-V-v}
V_t=\frac{V_x^2V_{xxx}}{V_{xx}^3}
+\frac{V_x}{V_{xx}}
-\frac{8}{9}x,
\end{gather}
where 
\begin{gather*}
V_x=-\frac{1}{27}\frac{1}{\tilde v_x^3}.
\end{gather*}
The first-order potentialisation of (\ref{Case-II-b-v}), with $v_x=W$, gives
\begin{gather}
\label{Case-II-b-v-W}
W_t=
\frac{W_{xxx}}{W_x^3}
-3\frac{W_{xx}^2}{W_x^4}
-3\frac{W_{xx}}{WW_x^2}
-\frac{3}{W^2};
\end{gather}
and the first-order potentialisation of (\ref{Case-II-b-v-vt}), with $\tilde v_x=\tilde W$, gives
\begin{gather}
\label{Case-II-b-vt-Wt}
\tilde W_t=
\frac{\tilde W^6\tilde W_{xxx}}{\tilde W_x^3}
-3\frac{\tilde W^6\tilde W_{xx}^2}{\tilde W_x^4}
+9\frac{\tilde W^5\tilde W_{xx}}{\tilde W_x^2}
-15\tilde W^4.
\end{gather}
Another first-order potentialisation for  (\ref{Case-II-b-v-vt}) is 
\begin{gather}
\tilde V_t=-\frac{\tilde V^{3/2}\tilde V_{xxx}}{\tilde V_x^3}+3\frac{\tilde V^{3/2}\tilde V^2_{xx}}{\tilde V_x^4},
\end{gather}
where we have applied the following integrating factor of (\ref{Case-II-b-v-vt}):
\begin{gather}
\Lambda[\tilde v]=-\frac{3}{2}\frac{\tilde v_{xx}}{\tilde v_x^4}
\end{gather}
and the corresponding conserved current
\begin{gather}
\Phi^t[\tilde v]=\frac{1}{4}\frac{1}{\tilde v_x^2}
\end{gather}
for the first-order potentialisation $\tilde V_x=D_x(\Phi^t)$, so 
$
\tilde V=4^{-1}\tilde v_x^{-2}.
$
Moreover, the first-order potentialisation of (\ref{Case-II-b-V-v}), with $V_x=q$, gives
\begin{gather}
\label{Case-II-b-V-q}
q_t=\frac{q^2q_{xxx}}{q_x^3}
-3\frac{q^2q_{xx}^2}{q_x^4}
+\frac{qq_{xx}}{q_x^2}
+\frac{1}{9}.
\end{gather}
Diagram 1 displays this multi-potentialisation of equation (\ref{Case-II-eq-lam-0-1-N}).

\item[c)] Using the conserved current $^{II}\Phi^t_{1,3}$, given by (\ref{Phi-t-II-lam-1-0}), we find that equation (\ref{Case-II-eq-lam-1-0-N})
viz.
\begin{gather*}
\boxed{\vphantom{\frac{DA}{DB}}
u_t=\lambda_1^3\frac{u_{xx}^3}{u_{xxx}^2}
}
\end{gather*}
admits the zero-order potentialisation
\begin{gather}
\label{Case-II-IF-3-vc}
v_t=-\frac{\lambda_1^3}{4}\left(\frac{v_x^3v_{xxx}}{v_{xx}^3}-3\frac{v_x^2}{v_{xx}}\right),
\end{gather}
where 
\begin{gather*}
v_x=k\lambda_1^{-3/2}u_{xx}^{1/2}
\end{gather*}
for any non-zero constant $k$.
We now consider $\lambda_1=-1$, so equation (\ref{Case-II-eq-lam-1-0-N}) takes the form
\begin{gather}
\label{3.38NEW}
\boxed{\vphantom{\frac{DA}{DB}}
u_t=-\frac{u_{xx}^3}{u_{xxx}^2}
}
\end{gather}
Turning to the multi-potentialisation of (\ref{3.38NEW}), we find the zero-order potentialisation of 
(\ref{Case-II-IF-3-vc}) in
\begin{gather}
\label{3.51}
\tilde v_t=2\frac{\tilde v_x^3\tilde v_{xxx}}{\tilde v_{xx}^3}-3\frac{\tilde v_x^2}{\tilde v_{xx}}\equiv 
2\frac{\tilde v_x^4}{\tilde v_{xx}^3}\, S[\tilde v],
\end{gather}
where 
\begin{gather}
\tilde v_x=v_{x}^2
\end{gather}
and $S$ is the Schwarzian (\ref{Schwarzian}). Furthermore (\ref{3.51}) admits the zero-order potentialisation
\begin{gather}
\label{Case-II-IF-3-Vc}
V_t=\frac{V_{xxx}}{V_{xx}^3}-3\cdot 2^{-2/3}\frac{1}{V_{xx}}-2^{-1/3}x,
\end{gather}
where 
\begin{gather}
V_x=2^{-1/3}\ln(\tilde v_x).
\end{gather}
Equation (\ref{Case-II-IF-3-Vc}) does not admit a zero-order potentialisation. 

\smallskip

\noindent
Turning to the first-order potentialisations of (\ref{3.38NEW}), 
we find the following:

\begin{itemize}
\item[$\bullet$]
With $v_x=w$, equation (\ref{Case-II-IF-3-vc}) takes the form
\begin{gather}
\label{Case-IIc-1st-Ord-w}
w_t=\frac{1}{4}\frac{w^3w_{xxx}}{w_x^3}-\frac{3}{4}\frac{w^3w_{xx}^2}{w_x^4}+\frac{3}{2}\frac{w^2w_{xx}}{w_x^2}-\frac{3}{2}w;
\end{gather}
\item[$\bullet$]
With $\tilde v_x=\tilde w$, equation (\ref{3.51})
takes the form
\begin{gather}
\label{Case-IIc-1st-Ord-wt}
\tilde w_t=2\frac{\tilde w^3\tilde w_{xxx}}{\tilde w_x^3}-6\frac{\tilde w^3\tilde w_{xx}^2}{\tilde w_x^4}+9\frac{\tilde w^2\tilde w_{xx}}{\tilde w_x^2}-6\tilde w;
\end{gather}
\item[$\bullet$]
With $V_x=W$, equation (\ref{Case-II-IF-3-Vc}) take the form
\begin{gather}
\label{Un-Pot-Eqi}
W_t=\frac{W_{xxx}}{W_x^3}-3\frac{W_{xx}^2}{W_x^4}+3\cdot 2^{-2/3}\frac{W_{xx}}{W_x^2}-2^{-1/3}.
\end{gather}
\end{itemize}
We find that equation (\ref{Case-II-IF-3-Vc}) admits also a second-order potentialisation which is obtained from 
equation (\ref{Un-Pot-Eqi}) with $W_x=p$, namely
\begin{gather}
\label{Un-Potj}
p_t=\frac{p_{xxx}}{p^3}
-9\,\frac{p_x p_{xx}}{p^4}
+12\,\frac{p_x^3}{p^5}
+3\,\cdot2^{-2/3}\,\frac{p_{xx}}{p^2}
-3\,\cdot2^{1/3}\,\frac{p_x^2}{p^3}.
\end{gather}
Note that (\ref{Un-Potj}) is a spacial case of one of the integrable equations list in \cite{MSS}, 
namely equation (4.1.34).
Interestingly, equation (\ref{Un-Potj}) admits only one integrating factor, namely $\Lambda[p]=1$, which leads to the following zero-order potentialisation of (\ref{Un-Potj}):
\begin{gather}
\label{Case-II-q-shabat-eq}
q_t=\frac{q_{xxx}}{q_x^3}-3\frac{q_{xx}^2}{q_x^4}+3\cdot 2^{-2/3}\,\frac{q_{xx}}{q_x^2},
\end{gather}
where $q_x=p$. Diagram 2 displays this multi-potentialisation of equation (\ref{3.38NEW}).

\smallskip

Furthermore we find that (\ref{Case-II-q-shabat-eq}) admits three integrating factors of order zero (none of order two or order four), namely 
\begin{gather}
\Lambda_1[q]=1,\qquad  \Lambda_2[q]=q, \ \ \mbox{and}\ \ \Lambda_3[q]=\exp\left(3\cdot 2^{-2/3} q\right), 
\end{gather}
which give the following potentialisations for (\ref{Case-II-q-shabat-eq}):
\begin{itemize}
\item[$\bullet$]
For $\Lambda_1[q]=1$ we obtain the following zero-order potentialisation of (\ref{Case-II-q-shabat-eq}):
\begin{gather}
\label{Case-II-Q1-shabat-eq}
Q_{1,t}=\frac{Q_{1,xxx}}{Q_{1,xx}^3}-3\cdot 2^{-2/3}\frac{1}{Q_{1,xx}},
\end{gather}
where 
\begin{gather}
Q_{1,x}=q.
\end{gather}
Obviously the first order potentialisation of (\ref{Case-II-Q1-shabat-eq}) leads back to 
(\ref{Case-II-q-shabat-eq}).

\item[$\bullet$]
For $\Lambda_1[q]=q$ we obtain the following zero-order potentialisation of (\ref{Case-II-q-shabat-eq}):
\begin{gather}
\label{Case-II-Q2-shabat-eq}
Q_{2,t}=\frac{Q_{2,x}^{3/2}Q_{2,xxx}}{Q_{2,xx}^3}
-3\cdot 2^{-2/3}\,\frac{Q_{2,x}}{Q_{2,xx}}
+3\cdot 2^{-5/3}\, x,
\end{gather}
where 
\begin{gather}
Q_{2,x}=\frac{q^2}{4}.
\end{gather}
The corresponding first-order potentialisation of (\ref{Case-II-q-shabat-eq}) is then
\begin{gather}
\tilde Q_{2,t}=\frac{\tilde Q_2^{3/2}\tilde Q_{2,xxx}}{\tilde Q_{2,x}^3}
-3\frac{\tilde Q_2^{3/2}\tilde Q_{2,xx}^2}{\tilde Q_{2,x}^4}
+\frac{3}{2} \frac{\left(\tilde Q_2^{1/2}
+2^{1/3}\tilde Q_2\right)\tilde Q_{2,xx}}{\tilde Q_{2,x}^2}\nn\\[0.3cm]
\label{Case-II-Q2t-shabat-eq}
\qquad
-3\cdot 2^{-5/3},
\end{gather}
where 
\begin{gather}
Q_{2,x}=\tilde Q_2.
\end{gather}

\item[$\bullet$]
For $\Lambda_1[q]=\exp\left(3\cdot 2^{-2/3}q\right)$ we obtain the following zero-order potentialisation of (\ref{Case-II-q-shabat-eq}):
\begin{gather}
\label{Case-II-Q3-shabat-eq}
Q_{3,t}=\frac{27}{4}\frac{Q_{3,x}^{3}Q_{3,xxx}}{Q_{3,xx}^3}
-\frac{27}{4}\frac{Q_{3,x}^2}{Q_{3,xx}},
\end{gather}
where 
\begin{gather}
Q_{3,x}=\exp\left(3\cdot 2^{-2/3}q\right).
\end{gather}
The corresponding first-order potentialisation of (\ref{Case-II-q-shabat-eq}) is then
\begin{gather}
\label{Case-II-Q3t-shabat-eq}
\tilde Q_{3,t}=\frac{27}{4}\frac{\tilde Q_3^{3}\tilde Q_{3,xxx}}{\tilde Q_{3,x}^3}
-\frac{81}{4}\frac{\tilde Q_3^{3}\tilde Q_{3,xx}^2}{\tilde Q_{3,x}^4}
+27\frac{\tilde Q_3^2\tilde Q_{3,xx}}{\tilde Q_{3,x}^3},
\end{gather}
where 
\begin{gather}
Q_{3,x}=\tilde Q_3.
\end{gather}

\end{itemize}

\noindent
Diagram 3 displays the potentialisation of equation (\ref{Case-II-q-shabat-eq}).

\end{itemize}

}


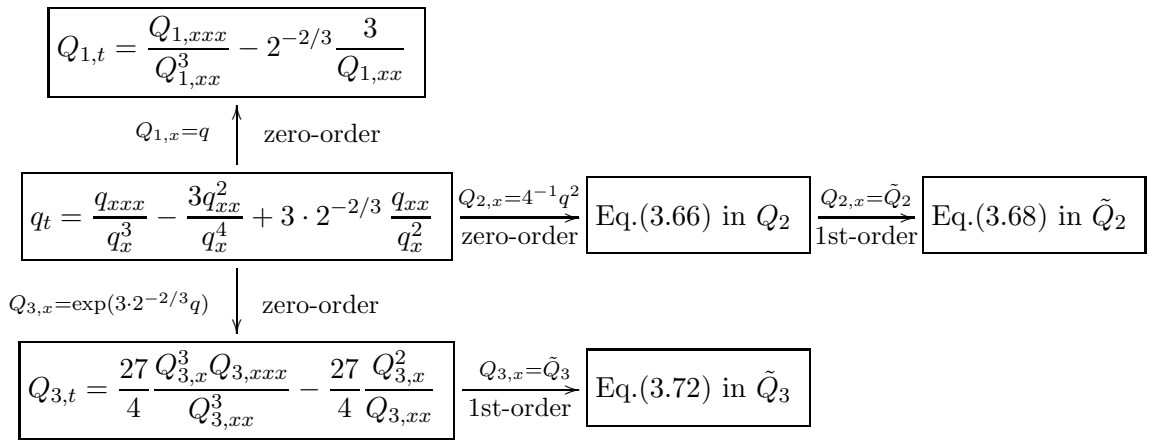
\strut\hfill


\begin{center}

\qquad\qquad\qquad
  \xymatrix{
&\qquad\qquad\mbox{{\bf Diagram 1}}\qquad\qquad & \qquad\qquad\qquad\qquad\quad\\
&
\boxed{
\vphantom{\frac{DA}{DB}}
\ u_t=\frac{u_{xx}^6}{u_{xxx}^2}\ 
}
\ar[d]^{\quad v_x=2^{-1/3}u_{xx}^{-1} }_{{\small\mbox{zero-order}}\quad }
\\
&
\boxed{
\vphantom{\frac{DA}{DB}}
v_t=\frac{v_{xxx}}{v_{xx}^3}+\frac{3}{v_xv_{xx}}
}
\ar[r]^{v_x=W}_{{\small\mbox{1st-order}}}
\ar[d]^{\quad \tilde v_{x}=-v_x^{-1}}_{{\small \mbox{zero-order}\quad }} & 
\boxed{
\vphantom{\frac{DA}{DB}}
\mbox{Eq.} (\ref{Case-II-b-v-W})\ \mbox{in}\  W
}
\\
\boxed{
\vphantom{\frac{DA}{DB}}
\tilde V_t=-\frac{\tilde V^{3/2}\tilde V_{xxx}}{\tilde V_x^3}
+3\frac{\tilde V^{3/2}\tilde V_{xx}^2}{\tilde V_x^4}\ 
}
&\ar[l]^{{\qquad\quad\small\mbox{1st-order}}}_{\qquad\quad \tilde V=4^{-1}\tilde v_{x}^{-2}}
\boxed{
\vphantom{\frac{DA}{DB}}
\tilde v_t=\frac{\tilde v_x^6\tilde v_{xxx}}{\tilde v_{xx}^3}-3\frac{\tilde v_x^5}{\tilde v_{xx}}
}
\ar[r]^{\quad \tilde v_x=\tilde W}_{{\small\mbox{\quad 1st-order}}}\ar[d]^{\quad V_x=-27^{-1}\tilde v_x^{-3}}_{{\small\mbox{zero-order}}\quad }  &
\boxed{
\vphantom{\frac{DA}{DB}}
\mbox{Eq.} (\ref{Case-II-b-vt-Wt})\ \mbox{in}\  \tilde W
}
\\
&
\boxed{
\vphantom{\frac{DA}{DB}}
V_t=\frac{V_x^2V_{xxx}}{V_{xx}^3}+\frac{V_x}{V_{xx}}-\frac{8}{9}x
}
\ar[r]^{\qquad V_x=q}_{{\small\mbox{\qquad\ \, 1st-order}}} &
\boxed{
\vphantom{\frac{DA}{DB}}
\mbox{Eq.} (\ref{Case-II-b-V-q})\ \mbox{in}\  q\ 
}
}

\end{center}

\strut\vfill

\pagebreak

\noindent
\begin{center}
\qquad\qquad\qquad
  \xymatrix{
  \mbox{{\bf \qquad\qquad\qquad   Diagram 2\qquad\qquad\qquad }}&\qquad\qquad& \qquad\qquad\qquad\qquad\qquad\quad \\
\boxed{
\vphantom{\frac{DA}{DB}}
\ u_t=-\frac{u_{xx}^3}{u_{xxx}^2}\ 
}
\ar[d]^{\quad v_x=-i (u_{xx})^{1/2} }_{{\small \mbox{zero-order}}\quad }
\\
\boxed{
\vphantom{\frac{DA}{DB}}
v_t=\frac{1}{4}\frac{v_x^3v_{xxx}}{v_{xx}^3}-\frac{3}{4}\frac{v_x^2}{v_{xx}}
%
}
\ar[r]^{v_x=w}_{{\small\mbox{1st-order}}}
\ar[d]^{\quad \tilde v_{x}=v_x^2}_{{\small\mbox{zero-order}}\quad } & 
\boxed{
\vphantom{\frac{DA}{DB}}
\mbox{Eq.} (\ref{Case-IIc-1st-Ord-w})\ \mbox{in}\  w
}
\\
\boxed{
\vphantom{\frac{DA}{DB}}
\tilde v_t=2\frac{\tilde v_x^3\tilde v_{xxx}}{\tilde v_{xx}^3}-3\frac{\tilde v_x^2}{\tilde v_{xx}}
}
\ar[r]^{\tilde v_x=\tilde w}_{{\small\mbox{1st-order}}}\ar[d]^{\quad V_{x}=2^{-1/3}\ln(\tilde v_x)}_{{\small\mbox{zero-order}}\quad } & 
\boxed{
\vphantom{\frac{DA}{DB}}
\mbox{Eq.} (\ref{Case-IIc-1st-Ord-wt})\ \mbox{in}\  \tilde w
}
\\
\boxed{
\vphantom{\frac{DA}{DB}}
V_t=\frac{V_{xxx}}{V_{xx}^3}-3\cdot 2^{-2/3}\frac{1}{V_{xx}}-2^{-1/3}x
}
\ar[r]^{\qquad\qquad \tilde V_x=W}_{\qquad\qquad {\small\mbox{1st-order}}}
& 
\boxed{
\vphantom{\frac{DA}{DB}}
\mbox{Eq.} (\ref{Un-Pot-Eqi})\ \mbox{in}\  W
}\ar[r]^{\tilde W_x=p}_{{\small\mbox{2nd-order}}} 
& \boxed{
\vphantom{\frac{DA}{DB}}
\mbox{Eq.} (\ref{Un-Potj})\ \mbox{in}\  p}\ar[d]^{\quad q_x=p}_{{\small\mbox{zero-order\quad }}}
\\
& & 
 \boxed{
\vphantom{\frac{DA}{DB}}
\mbox{Eq.} (\ref{Case-II-q-shabat-eq})\ \mbox{in}\  q}
}
\end{center}

\strut\hfill


\begin{displaymath}
\xymatrix{
 \mbox{{\bf\qquad\qquad\qquad Diagram 3 \qquad\qquad \quad\ }}&\qquad\qquad\qquad&\qquad\qquad\qquad  \qquad\qquad\\
\boxed{
\vphantom{\frac{DA}{DB}}
Q_{1,t}=\frac{Q_{1,xxx}}{Q_{1,xx}^3}-2^{-2/3}\frac{3}{Q_{1,xx}}\
}
\\
\boxed{
\vphantom{\frac{DA}{DB}}
q_t=\frac{q_{xxx}}{q_x^3}-\frac{3q_{xx}^2}{q_x^4}+3\cdot 2^{-2/3}\,\frac{q_{xx}}{q_x^2}\
}
\ar[r]^{\qquad\quad \ \,\,Q_{2,x}=4^{-1}q^2}_{\qquad\quad\ \,\,{\small\mbox{zero-order}}}\ar[u]^{Q_{1,x}=q\ \ }_{\ \ {\small\mbox{zero-order}}}\ar[d]_{Q_{3,x}=\exp(3\cdot 2^{-2/3} q)\ \ }^{{\small \mbox{\ \ zero-order}}} & 
\boxed{
\vphantom{\frac{DA}{DB}}
\mbox{Eq.(\ref{Case-II-Q2-shabat-eq}) in } Q_2\
}
\ar[r]^{Q_{2,x}=\tilde Q_{2}}_{{\small \mbox{1st-order}}} &
\boxed{
\vphantom{\frac{DA}{DB}}
\mbox{Eq.(\ref{Case-II-Q2t-shabat-eq}) in } \tilde Q_2\
}
\\
\boxed{
\vphantom{\frac{DA}{DB}}
Q_{3,t}=\frac{27}{4}\frac{Q_{3,x}^{3}Q_{3,xxx}}{Q_{3,xx}^3}
-\frac{27}{4}\frac{Q_{3,x}^2}{Q_{3,xx}}\ 
}
\ar[r]^{\qquad\qquad Q_{3,x}=\tilde Q_3}_{\qquad\quad\ \,\,\mbox{{\small 1st-order}}} & 
\boxed{
\vphantom{\frac{DA}{DB}}
\mbox{Eq.(\ref{Case-II-Q3t-shabat-eq}) in } \tilde Q_3\
}
\\
}
\end{displaymath}

\strut\hfill

\noindent
{\bf Subcase II.2:} We consider the integrating factor (\ref{Case-II-IF-1}) for equation (\ref{Case-II-eq}), viz.
\begin{gather*}
u_t=\frac{u_{xx}^3\left(\lambda_1+\lambda_2 u_{xx}\right)^3}{u_{xxx}^2}.
\end{gather*}

\strut\hfill

\noindent
\underline{Let $\lambda_1\neq 0$ and $\lambda_2\neq 0$:}
The integrating factor (\ref{Case-II-IF-1}) then leads to the following conserved current and flux 
for (\ref{Case-II-eq}):
\begin{subequations}
\begin{gather}
\label{Phi-t-II-lam-2-1}
^{II}\Phi^t_{2,1}[u]=\frac{2^{-1/3}}{\lambda_1}\ln \left(\lambda_2+\frac{\lambda_1}{u_{xx}}\right)
\\[0.3cm]
^{II}\Phi^x_{2,1}[x,u]=
-2^{2/3}\frac{(\lambda_1 +\lambda_2 u_{xx})^2u_{xx}^2u_{4x}}{u_{xxx}^3}
+5\cdot 2^{-1/3}\frac{(\lambda_1+\lambda_2 u_{xx})(\lambda_1+2\lambda_2 u_{xx})u_{xx}}{u_{xxx}}
\nn\\[0.3cm]
\label{Phi-x-II-lam-2-1}
\qquad
+2^{-1/3}\lambda_1^2\,x.
\end{gather}
\end{subequations}

\noindent
This leads to 

\strut\hfill

\noindent
{\bf Potentialisation II.2} {\it 

\smallskip

\noindent
Using the conserved current $^{II}\Phi^t_{2,1}$, given by (\ref{Phi-t-II-lam-2-1}), we find that equation 
(\ref{Case-II-eq}) viz.
\begin{gather*}
\boxed{\vphantom{\frac{DA}{DB}}
u_t=\frac{u_{xx}^3\left(\lambda_1+\lambda_2 u_{xx}\right)^3}{u_{xxx}^2}
}
\end{gather*}
admits the zero-order potentialisation
\begin{subequations}
\begin{gather}
\label{Case-II-IF-1-va}
v_t=\frac{v_{xxx}}{v_{xx}^3}
+\lambda_1 3\cdot 2^{-2/3}\frac{e^{2^{1/3}\lambda_1 v_x}+\lambda_2}{\left(e^{2^{1/3}\lambda_1 v_x}-\lambda_2\right)v_{xx}}
-2^{-1/3}\lambda_1^2\,x,
\end{gather}
where
\begin{gather}
v_x=\frac{2^{-1/3}}{\lambda_1}\ln\left(\frac{\lambda_1+\lambda_2 u_{xx}}{u_{xx}}\right).
\end{gather}
The first-order potentialisation of (\ref{Case-II-eq})  with $w_x=D_x(^{II}\Phi_{2, 1}^t)$ i.e.
$w=v_x$, is then
\begin{gather}
w_t=\frac{w_{xxx}}{w_x^3}-3\frac{w_{xx}^2}{w_x^4}
- 3\cdot 2^{-2/3}\lambda_1 \frac{\left(e^{2^{1/3}\lambda_1  w}+\lambda_2\right)w_{xx} }
{\left(e^{2^{1/3}\lambda_1 w}-\lambda_2\right)w_{x}^2}\nn\\[0.3cm]
\qquad
- 3\cdot 2^{2/3}\lambda_1^2\lambda_2
\frac{e^{2^{1/3}\lambda_1 w}}{(e^{2^{1/3}\lambda_1 w}-\lambda_2)^2}-2^{-1/3}\lambda_1^2.
\end{gather}
\end{subequations}

}

\strut\hfill

\noindent
We remark that the case $\lambda_1=1$ and $\lambda_2=0$, as well as the case $\lambda_1=0$ and $\lambda_2=1$, do not lead to different equations than those already listed in Potentialisation II.1.


\strut\hfill

\noindent
{\bf Subcase II.3:} We consider the integrating factor (\ref{Case-II-IF-2})
for equation (\ref{Case-II-eq}), viz.
\begin{gather*}
u_t=\frac{u_{xx}^3\left(\lambda_1+\lambda_2 u_{xx}\right)^3}{u_{xxx}^2}.
\end{gather*}

\strut\hfill

\noindent
\underline{Let $\lambda_1\neq 0$ and $\lambda_2\neq 0$:}  The integrating factor (\ref{Case-II-IF-2}) then leads to the following conserved current and flux for (\ref{Case-II-eq}):
\begin{subequations}
\begin{gather}
\label{Phi-t-II-lam-3-1}
^{II}\Phi^t_{3,1}[u]=\frac{1}{\lambda_1^2}\,
u_{xx}\ln\left(
\frac{u_{xx}}{\lambda_1+\lambda_2u_{xx}}\right)
+\frac{1}{\lambda_1\lambda_2}
\\[0.3cm]
\label{Phi-x-II-lam-3-1}
^{II}\Phi^x_{2,1}[u]=\frac{2}{\lambda_1^2}
\frac{(\lambda_1+\lambda_2 u_{xx})^2u_{xx}^3u_{4x}}{u_{xxx}^3}
\left[
\lambda_1+(\lambda_1+\lambda_2 u_{xx})\ln
\left(\frac{u_{xx}}{\lambda_1+\lambda_2 u_{xx}}\right)\right]\nn\\[0.3cm]
\qquad
-\frac{3}{\lambda_1^2}
\frac{u_{xx}^2}{u_{xxx}}
(\lambda_1+2\lambda_2u_{xx})(\lambda_1+\lambda_2 u_{xx})^2
\ln\left(
\frac{u_{xx}}{\lambda_1+\lambda_2u_{xx}}\right)\nn\\[0.3cm]
\qquad
-\frac{1}{\lambda_1}
\frac{u_{xx}^2}{u_{xxx}}
(\lambda_1+6\lambda_2 u_{xx})(\lambda_1+\lambda_2 u_{xx})
-\lambda_1 u_x.
\end{gather}
\end{subequations}

\strut\hfill

\noindent
\underline{Let $\lambda_1=0$ and $\lambda_2= 1$:}  In this case the integrating factor
 (\ref{Case-II-IF-2}) is identical to the integrating factor (\ref{Case-II-IF-1}), which has already been described in Subcase II.2.

\strut\hfill

\noindent
\underline{Let $\lambda_1=1$ and $\lambda_2= 0$:}  
Equation (\ref{Case-II-eq}) then takes the form (\ref{Case-II-eq-lam-1-0-N}) with $\lambda_1=1$, viz.
\begin{gather*}
u_t=\frac{u_{xx}^3}{u_{xxx}^2}
\end{gather*}
and the integrating factor (\ref{Case-II-IF-2}) leads to the following conserved current and flux for 
(\ref{Case-II-eq-lam-1-0-N}) with $\lambda_1=1$:
\begin{subequations}
\begin{gather}
\label{Phi-t-II-3-lam-1-0}
^{II}\Phi^t_{3,3}[u]=u_{xx}\ln(u_{xx})-u_{xx}
\\[0.3cm]
\label{Phi-x-II-3--lam-1-0}
^{II}\Phi^x_{3,3}[u]=
\frac{2u_{xx}^3u_{4x}}{u_{xxx}^3}\ln(u_{xx})
-\frac{3u_{xx}^2}{u_{xxx}}\ln(u_{xx})
+\frac{2u_{xx}^2}{u_{xxx}}
-u_x.
\end{gather}
\end{subequations}

\noindent
This leads to 

\strut\hfill

\noindent
{\bf Potentialisation II.3} {\it 

\begin{itemize}

\item[a)]

Using the conserved current $^{II}\Phi^t_{3,1}$, given by (\ref{Phi-t-II-lam-3-1}), we find that equation 
(\ref{Case-II-eq}) viz.
\begin{gather*}
\boxed{\vphantom{\frac{DA}{DB}}
u_t=\frac{u_{xx}^3\left(\lambda_1+\lambda_2 u_{xx}\right)^3}{u_{xxx}^2}
}
\end{gather*}
admits no zero-order potentialisation. Equation (\ref{Case-II-eq}) does however admit a first-order potentialisation with
$w_x=D_x(^{II}\Phi^t_{3,1})$, i.e. 
\begin{gather}
\label{Pot-II-1-w}
w(x,t)=\frac{u_{xx}}{\lambda_1^2}
\ln\left(
\frac{u_{xx}}{\lambda_1+\lambda_2 u_{xx}}\right)
+\frac{1}{\lambda_1\lambda_2},
\end{gather}
namely
\begin{gather}
w_t=-\frac{2}{\lambda_2}^3\frac{w_{xxx}}{w_x^3}
(\lambda_2^2u_{xx}w+\lambda_1\lambda_2 w-1)^3
+u_{xx}^3\left(
30\lambda_2^3 w+6\lambda_2^3\frac{w^3w_{xx}^2}{w_x^4}
-18\lambda_2^3\frac{w^2w_{xx}}{w_x^2}\right)\nn\\[0.3cm]
\qquad
+u_{xx}^2
\left(60\lambda_1\lambda_2^2 w
-18\lambda_2
-15\lambda_2\frac{w w_{xx}}{w_x^2}(3\lambda_1\lambda_2 w-2)
+18\lambda_2 \frac{w^2w_{xx}^2}{w_x^4}(\lambda_1\lambda_2w-1)\right)\nn\\[0.3cm]
\qquad
+\frac{6}{\lambda_2^3}\frac{w_{xx}^2}{w_x^4}(\lambda_1\lambda_2 w-1)^3
-\frac{9\lambda_1}{\lambda_2^2}\frac{w_{xx}}{w_x^2}(\lambda_1\lambda_2 w-1)^2
+6\lambda_1^3 w-\frac{6\lambda_1^2}{\lambda_2}\nn\\[0.3cm]
\qquad
+u_{xx}\left(36\lambda_1^2\lambda_2w-23\lambda_1+\frac{18}{\lambda_2}\frac{w w_{xx}^2}{w_x^4}(\lambda_1\lambda_2 w-1)^2\right.\nn\\[0.3cm]
\qquad
\left.
-\frac{12}{\lambda_2}\frac{w_{xx}}{w_x^2}(\lambda_1\lambda_2w-1)(3\lambda_1\lambda_2 w-1)\right),
\end{gather}
where $u_{xx}$ needs to be solved algebraically in terms of $w$ from (\ref{Pot-II-1-w}), 
which is possible by the Lambert function.

\item[b)] Using the conserved current $^{II}\Phi_{3,3}^t$ given by (\ref{Phi-t-II-3-lam-1-0}), we find that equation (\ref{Case-II-eq-lam-1-0-N}) with $\lambda_1=1$, viz.
\begin{gather*}
\boxed{\vphantom{\frac{DA}{DB}}
u_t=\frac{u_{xx}^3}{u_{xxx}^2}
}
\end{gather*}
 does not admit a zero-order potentialisation but it does admit a first-order potentialisation with
$w_x=D_x(^{II}\Phi^t_{3,3})$, i.e. 
\begin{gather}
\label{377}
w=u_{xx}\ln(u_{xx})-u_{xx},
\end{gather}
namely
\begin{gather}
w_t=-2\frac{w_{xxx}}{w_x^3}(u_{xx}+w)^3
+6\frac{w_{xx}^2}{w_x^4}(u_{xx}+w)^3
-3\frac{w_{xx}}{w_x^2}(u_{xx}+w)(5u_{xx}+3w)\nn\\[0.3cm]
\qquad
+6w+13u_{xx},
\end{gather}
where $u_{xx}$ needs to be solved algebraically from the relation (\ref{377}) in terms of $w$ 
by the Lambert function.
\end{itemize}

}

\strut\hfill

\noindent
{\bf Subcase II.4:} We consider the integrating factor (\ref{Case-II-IF-5})
for equation (\ref{Case-II-eq}), viz.
\begin{gather*}
u_t=\frac{u_{xx}^3\left(\lambda_1+\lambda_2 u_{xx}\right)^3}{u_{xxx}^2}.
\end{gather*}

\strut\hfill

\noindent
\underline{Let $\lambda_1\neq 0$ and $\lambda_2\neq 0$:}  The integrating factor (\ref{Case-II-IF-5}) then leads to the following conserved current and flux for (\ref{Case-II-eq}):
\begin{subequations}
\begin{gather}
\label{Phi-t-II-lam-4-1}
^{II}\Phi^t_{4,1}[u]=\frac{2^{-1/3}}{\lambda_1^2}(\lambda_1+\lambda_2 u_{xx})
\ln\left(\frac{u_{xx}}{\lambda_1+\lambda_2 u_{xx}}\right)+\frac{2^{-1/3}}{\lambda_1}
\\[0.3cm]
\label{Phi-x-II-lam-4-1}
^{II}\Phi^x_{4,1}[x,u]=2^{-2/3}\frac{1}{\lambda_1^2}
\frac{u_{xx}^2u_{4x}}{u_{xxx}^2}(\lambda_1+\lambda_2 u_{xx})^3\left[
\lambda_2u_{xx}\ln\left(\frac{u_{xx}}{\lambda_1+\lambda_2 u_{xx}}\right)+\lambda_1\right]\nn\\[0.3cm]
\qquad
-3\cdot 2^{-1/3}\frac{\lambda_2}{\lambda_1^2}
\frac{u_{xx}^2}{u_{xxx}}(\lambda_1+2\lambda_2 u_{xx})(\lambda_1+\lambda_2 u_{xx})^2  
\ln\left(\frac{u_{xx}}{\lambda_1+\lambda_2 u_{xx}}\right)\nn\\[0.3cm]
\qquad
-2^{-1/3}\frac{1}{\lambda_1}\frac{u_{xx}}{u_{xxx}}(5\lambda_1+6\lambda_2 u_{xx})
(\lambda_1+\lambda_2 u_{xx})^2
-2^{-1/3}\lambda_1\lambda_2 u_x
-2^{-1/3}\lambda_1^2 x.
\end{gather}
\end{subequations}

\strut\hfill

\noindent
\underline{Let $\lambda_1=0$ and $\lambda_2=1$:}  The integrating factor (\ref{Case-II-IF-5}) is then identical  to integrating factor (\ref{Case-II-IF-3}) described in Subcase II.1.

\strut\hfill

\noindent
\underline{Let $\lambda_1=1$ and $\lambda_2=0$:}  The integrating factor (\ref{Case-II-IF-5}) then leads to the following conserved current and flux for  (\ref{Case-II-eq-lam-1-0-N}) with $\lambda_1=1$, viz
\begin{gather*}
u_t=\frac{u_{xx}^3}{u_{xxx}^2},
\end{gather*}
namely
\begin{subequations}
\begin{gather}
\label{Phi-t-II-lam-4-3}
^{II}\Phi^t_{4,3}[u]=-2^{-1/3}\ln (u_{xx})\\[0.3cm]
\label{Phi-x-II-lam-4-3}
^{II}\Phi^x_{4,3}[x,u]=-2^{2/3}\left(\frac{u_{xx}^2u_{4x}}{u_{xxx}^3}
-\frac{5}{2}\frac{u_{xx}}{u_{xxx}}
-\frac{1}{2}x\right).
\end{gather}
\end{subequations}

\noindent
This leads to 

\strut\hfill

\noindent
{\bf Potentialisation II.4} {\it

\begin{itemize}

\item[a)]
Using the conserved current $^{II}\Phi^t_{4,1}$, given by (\ref{Phi-t-II-lam-4-1}), we find that equation 
(\ref{Case-II-eq}) viz.
\begin{gather*}
\boxed{\vphantom{\frac{DA}{DB}}
u_t=\frac{u_{xx}^3\left(\lambda_1+\lambda_2 u_{xx}\right)^3}{u_{xxx}^2}
}
\end{gather*}
admits no zero-order potentialisation. Equation (\ref{Case-II-eq}) does however admit a first-order potentialisation with
$w_x=D_x(^{II}\Phi^t_{4,1})$, i.e. 
\begin{gather}
\label{Pot-II-3-w}
w(x,t)=\frac{2^{-1/3}}{\lambda_1^2}(\lambda_1+\lambda_2 u_{xx})
\ln\left(\frac{u_{xx}}{\lambda_1+\lambda_2 u_{xx}}\right)+\frac{2^{-1/3}}{\lambda_1}
\end{gather}
namely
\begin{gather}
w_t=
-\frac{w_{xxx}}{w_x^3}
\frac{(2^{1/3}\lambda_2w u_{xx}+1)^3}{(\lambda_2 u_{xx}+\lambda_1)^2}
+3\frac{w_{xx}^2}{w_x^4}
\frac{(2^{1/3}\lambda_2w u_{xx}+1)^3}{(\lambda_2u_{xx}+\lambda_1)^2}\nn\\[0.3cm]
-\frac{9\lambda_2^2w_{xx}w^2}{w_x^2}
\frac{(2\lambda_2 u_{xx}+\lambda_1)u_{xx}^2}{(\lambda_2u_{xx}+\lambda_1)^2}
+
\frac{3\cdot 2^{2/3}\lambda_2w_{xx}w}{\lambda_1w_x^2}
\frac{(\lambda_2^2 u_{xx}^2-4\lambda_1\lambda_2u_{xx}-2\lambda_1^2)u_{xx}}{
(\lambda_2u_{xx}+\lambda_1)^2}\nn\\[0.3cm]
+\frac{3\cdot 2^{-2/3}\lambda_2w_{xx}}{\lambda_1^2w_x^2}
\frac{(4\lambda_2^2u_{xx}^2+11\lambda_1\lambda_2u_{xx}+4\lambda_1^2)u_{xx}}{
(\lambda_2u_{xx}+\lambda_1)^2}
-\frac{3\cdot 2^{-2/3}w_{xx}}{\lambda_1^2w_x^2}
(4\lambda_2u_{xx}+\lambda_1)\nn\\[0.3cm]
-\frac{3\lambda_2(2^{2/3}-2\lambda_1 w)}{\lambda_1}
\frac{(5\lambda_2^2u_{xx}^2+5\lambda_1\lambda_2u_{xx}+\lambda_1^2)u_{xx}}{(\lambda_2u_{xx}+\lambda_1)^2}\nn\\[0.3cm]
+\frac{2^{-1/3}}{\lambda_1}
\frac{(16\lambda_2^2u_{xx}^2+11\lambda_1\lambda_2 u_{xx}+\lambda_1^2)u_{xx}}{(\lambda_2u_{xx}+\lambda_1)^2},
\end{gather}
where $u_{xx}$ needs to be solved algebraically in $w$ from (\ref{Pot-II-3-w}) in terms of the Lambert function.

\item[b)] 
Using the conserved current $^{II}\Phi^t_{4,3}$, given by (\ref{Phi-t-II-lam-4-3}), we find that equation 
(\ref{Case-II-eq-lam-1-0-N}) with $\lambda_1=1$, viz.
\begin{gather*}
\boxed{\vphantom{\frac{DA}{DB}}
u_t=\frac{u_{xx}^3}{u_{xxx}^2}
}
\end{gather*}
admits the zero-order potentialisation
\begin{gather}
v_t=\frac{v_{xxx}}{v_{xx}^3}
+3\cdot 2^{-2/3}\frac{1}{v_{xx}}
-2^{-1/3}x,
\end{gather}
where 
\begin{gather}
v_x=-2^{-1/3}\ln (u_{xx}).
\end{gather}
Equation (\ref{Case-II-eq-lam-1-0-N}) also admits a first-order potentialisation with 
$w_x=D_x(^{II}\Phi^t_{4,3})$, i.e. $v_x=w$, namely
\begin{gather}
w_t=\frac{w_{xxx}}{w_x^3}-3\frac{w_{xx}^2}{w_x^4}
-3\cdot 2^{-2/3}\frac{w_{xx}}{w_x^2}
-2^{-1/3},
\end{gather}
and a second-order potentialisation with $W_x=D_x^2(^{II}\Phi^t_{4,3})$, i.e. $W=w_x$, 
namely
\begin{gather}
W_t=\frac{W_{xxx}}{W^3}
-9\frac{W_xW_{xx}}{W^4}
-3\cdot 2^{-2/3}\frac{W_{xx}}{W^2}
+12\frac{W_x^3}{W^5}
+3\cdot 2^{1/3}\frac{W_x^2}{W^3}.
\end{gather}

\end{itemize}
}

\strut\hfill

\noindent
{\bf Case III:} We consider (\ref{Case-III-eq}), viz.
\begin{gather*}
u_t=\frac{(\alpha u_x+\beta)^{11}}{\left[
\vphantom{\frac{DA}{DB}}
(\alpha u_x+\beta)u_{xxx}-3\alpha u_{xx}^2\right]^2},
\end{gather*}
where $\alpha$ and $\beta$ are arbitrary constants, not simultaneously zero.
Equation (\ref{Case-III-eq}) does not admit zero-order or second-order integrating factors, but the equation does admit the following three fourth-order integrating factors:
\begin{subequations}
\begin{gather}
\label{L-III-Gen-1}
^{III}\Lambda^1_1[u]=\frac{u_{xx}^2u_{4x}}{(\alpha u_x+\beta)^{11}}
+\frac{2u_{xx}u_{xxx}^2}{(\alpha u_x+\beta)^{11}}
-\frac{22\alpha u_{xx}^3u_{xxx}}{(\alpha u_x+\beta)^{12}}
+\frac{33\alpha^2u_{xx}^5}{(\alpha u_x+\beta)^{13}}\\[0.3cm]
\label{L-III-Gen-2}
^{III}\Lambda^1_2[u]=\frac{u_{xx}u_{4x}}{(\alpha u_x+\beta)^{8}}
+\frac{u_{xxx}^2}{(\alpha u_x+\beta)^{8}}
-\frac{16\alpha u_{xx}^2u_{xxx}}{(\alpha u_x+\beta)^{9}}
+\frac{24\alpha^2u_{xx}^4}{(\alpha u_x+\beta)^{10}}\\[0.3cm]
\label{L-III-Gen-3}
^{III}\Lambda^1_3[u]=\frac{u_{4x}}{(\alpha u_x+\beta)^{5}}
-\frac{10\alpha u_{xx}u_{xxx}}{(\alpha u_x+\beta)^{6}}
+\frac{15\alpha^2u_{xx}^3}{(\alpha u_x+\beta)^{7}}.
\end{gather}
\end{subequations}
Using (\ref{L-III-Gen-1}), (\ref{L-III-Gen-2}) and (\ref{L-III-Gen-3}), we obtain the following respective conserved currents for  (\ref{Case-III-eq}):
\begin{gather}
^{III}\Phi^t_{1,1}[u]=
\frac{u_{xx}^4}{(\alpha u_x+\beta)^{11}},\quad
^{III}\Phi^t_{1,2}[u]=
\frac{u_{xx}^3}{(\alpha u_x+\beta)^{8}},\nn\\[0.3cm]
\label{Ans-III-Gen}
\ \mbox{and}\ \ 
^{III}\Phi^t_{1,3}[u]=
\frac{u_{xx}^2}{(\alpha u_x+\beta)^{5}}.
\end{gather}
We find that all three conserved currents (\ref{Ans-III-Gen}) do not lead to a potentialisation of 
(\ref{Case-III-eq}), of any order.

\strut\hfill

\noindent
\underline{Let $\alpha=1$ and $\beta=0$:} Equation (\ref{Case-III-eq}) then takes the form
\begin{gather}
\label{Case-III-alpha-1-beta-0-eq}
u_t=\frac{u_x^{11}}{\left(
\vphantom{\frac{DA}{DB}}
u_xu_{xxx}-3u_{xx}^2\right)^2}.
\end{gather}
In addition to the integrating factors (\ref{L-III-Gen-1}), (\ref{L-III-Gen-2}) and (\ref{L-III-Gen-3}) with 
$\alpha=1$ and $\beta=0$, equation (\ref{Case-III-alpha-1-beta-0-eq}) also admits the following fourth-order  
integrating factors:
\begin{subequations}
\begin{gather}
^{III}\Lambda^2_1[u]=
\left(
\frac{uu_{xx}^2}{u_x^{11}}
+\frac{u_{xx}}{u_x^9}\right)u_{4x}
+\left(
\frac{1}{u_x^9}+\frac{2uu_{xx}}{u_x^{11}}\right)u_{xxx}^2
-\left(
\frac{22uu_{xx}^3}{u_x^{12}}+\frac{16 u_{xx}^2}{u_x^{10}}\right)u_{xxx}\nn\\[0.3cm]
\label{L-III-Gen-2-1}
\qquad
+\frac{33uu_{xx}^5}{u_x^{13}}
+\frac{24 u_{xx}^4}{u_x^{11}}\\[0.3cm]
^{III}\Lambda^2_2[u]=
\left(
\frac{uu_{xx}}{u_x^8}+\frac{1}{u_x^6}\right)u_{4x}
-\left(
\frac{16uu_{xx}^2}{u_x^9}
+\frac{10u_{xx}}{u_x^7}\right)u_{xxx}
+\frac{uu_{xxx}^2}{u_x^8}
+\frac{24uu_{xx}^4}{u_x^{10}}\nn\\[0.5cm]
\label{L-III-Gen-2-2}
\qquad
+\frac{15 u_{xx}^3}{u_x^8}.
\end{gather}
\end{subequations}
Using (\ref{L-III-Gen-2-1}) and (\ref{L-III-Gen-2-2}), we obtain the following respective conserved currents for (\ref{Case-III-alpha-1-beta-0-eq}):
\begin{gather}
\label{Ans-III-1-0}
^{III}\Phi^t_{2,1}[u]=
\frac{1}{12}\frac{uu_{xx}^4}{u_x^{11}}
+\frac{1}{6}\frac{u_{xx}^3}{u_x^9}\qquad \mbox{and}\qquad
^{III}\Phi^t_{2,2}[u]=
\frac{1}{6}\frac{uu_{xx}^3}{u_x^8}
+\frac{1}{2}\frac{u_{xx}^2}{u_x^6}.
\end{gather}
We find that both conserved currents (\ref{Ans-III-1-0}) do not lead to a potentialisation of 
(\ref{Case-III-alpha-1-beta-0-eq}), of any order.

\strut\hfill

\noindent
\underline{Let $\alpha=0$ and $\beta=1$:} Equation (\ref{Case-III-eq}) then takes the form
\begin{gather}
\label{Case-III-alpha-0-beta-1-eq}
u_t=\frac{1}{u_{xxx}^2},
\end{gather}
which admits the following three fourth-order integrating factors (no zero-order or second-order exist):
\begin{subequations}
\begin{gather}
\label{L-III-Gen-3-1}
^{III}\Lambda^3_1[u]=
u_{xx}^2u_{4x}+2u_{xx}u_{xxx}^2\\[0.3cm]
\label{L-III-Gen-3-2}
^{III}\Lambda^3_2[u]=
u_{xx}u_{4x}+u_{xxx}^2\\[0.3cm]
\label{L-III-Gen-3-3}
^{III}\Lambda^3_3[u]=
u_{4x}.
\end{gather}
\end{subequations}
Note that the integrating factors (\ref{L-III-Gen-3-1}),  (\ref{L-III-Gen-3-2}) and  (\ref{L-III-Gen-3-3}) are just
the integrating factors (\ref{L-III-Gen-1}), (\ref{L-III-Gen-2}) and (\ref{L-III-Gen-3}), respectively, with 
$\alpha=0$ and $\beta=1$.  No additional integrating factors up to order four than those listed here were obtained for equation  
(\ref{Case-III-alpha-0-beta-1-eq}). The corresponding conserved currents and fluxes are as follows:
\begin{subequations}
\begin{gather}
\label{Phi-t-III-lam-3-1}
^{III}\Phi^t_{3,1}[u]=2^{-1/3}\frac{1}{512}u_{xx}^4
\\[0.3cm]
\label{Phi-x-III-lam-3-1}
^{III}\Phi^x_{3,1}[u]=2^{2/3}\frac{1}{64}\left(
\frac{u_{xx}^3u_{4x}}{u_{xxx}^3}
+\frac{3u_{xx}^2}{u_{xxx}}
-3u_x\right)
\\[0.3cm]
\label{Phi-t-III-lam-3-2}
^{III}\Phi^t_{3,2}[u]=-\frac{1}{54}u_{xx}^2
\\[0.3cm]
\label{Phi-x-III-lam-3-2}
^{III}\Phi^x_{3,2}[x,u]=-\frac{1}{9}\left(\frac{u_{xx}^2u_{4x}}{u_{xxx}^3}
+\frac{2u_{xx}}{u_{xxx}}-2x\right)
\\[0.3cm]
\label{Phi-t-III-lam-3-3}
^{III}\Phi^t_{3,3}[u]=2^{-8/3}u_{xx}^2
\\[0.3cm]
\label{Phi-x-III-lam-3-3}
^{III}\Phi^x_{3,3}[u]=2^{-2/3}
\left(
\frac{u_{xx}u_{4x}}{u_{xxx}^3}
+\frac{1}{u_{xxx}}\right).
\end{gather}
\end{subequations}

\noindent
This leads to 

\strut\hfill

\noindent
{\bf Potentialisation III.1} {\it


\begin{itemize}

\item[a)]
Using the conserved current $^{III}\Phi^t_{3,1}$, given by (\ref{Phi-t-III-lam-3-1}), we find that equation 
(\ref{Case-III-alpha-0-beta-1-eq}) viz.
\begin{gather*}
\boxed{\vphantom{\frac{DA}{DB}}
u_t=\frac{1}{u_{xxx}^2}
}
\end{gather*}
admits no zero-order potentialisation. Equation (\ref{Case-III-alpha-0-beta-1-eq}) does however admit a first-order potentialisation with
$w_x=D_x(^{III}\Phi^t_{3,1})$, i.e.
\begin{gather}
w(x,t)=2^{-1/3}\frac{1}{512}u_{xx}^4,
\end{gather} 
namely
\begin{gather}
w_t=-\frac{w^{9/4}w_{xxx}}{w_x^3}
+3\frac{w^{9/4}w_{xx}^2}{w_x^4}
-\frac{9}{4}
\frac{w^{5/4}w_{xx}}{w_x^2}
+\frac{3}{8}w_x^{1/4}.
\end{gather}

\item[b)]
Using the conserved current $^{III}\Phi^t_{3,2}$, given by (\ref{Phi-t-III-lam-3-2}), we find that equation 
(\ref{Case-III-alpha-0-beta-1-eq}) viz.
\begin{gather*}
\boxed{\vphantom{\frac{DA}{DB}}
u_t=\frac{1}{u_{xxx}^2}
}
\end{gather*}
admits the zero-order potentialisation 
\begin{gather}
\label{Case-III-Multi-x}
v_t=\frac{v_x^2v_{xxx}}{v_{xx}^3}-\frac{2}{9}x,
\end{gather}
where 
\begin{gather}
v_x=-\frac{1}{54}u_{xx}^3.
\end{gather}
Equation (\ref{Case-III-alpha-0-beta-1-eq}) also admits a first-order potentialisation with
$w_x=D_x(^{III}\Phi^t_{3,2})$, i.e. $v_x=w$, namely
\begin{gather}
w_t=\frac{w^{2}w_{xxx}}{w_x^3}
-3\frac{w^{2}w_{xx}^2}{w_x^4}
+2
\frac{w w_{xx}}{w_x^2}
-\frac{2}{9}.
\end{gather}

\item[c)]
Using the conserved current $^{III}\Phi^t_{3,3}$, given by (\ref{Phi-t-III-lam-3-3}), we find that equation 
(\ref{Case-III-alpha-0-beta-1-eq}) viz.
\begin{gather*}
\boxed{\vphantom{\frac{DA}{DB}}
u_t=\frac{1}{u_{xxx}^2}
}
\end{gather*}
admits the zero-order potentialisation 
\begin{gather}
v_t=-\frac{v_x^{3/2}v_{xxx}}{v_{xx}^3},
\end{gather}
where 
\begin{gather}
v_x=2^{-8/3}u_{xx}^2.
\end{gather}
Equation (\ref{Case-III-alpha-0-beta-1-eq}) also admits a first-order potentialisation with
$w_x=D_x(^{III}\Phi^t_{3,3})$, i.e. $v_x=w$, namely
\begin{gather}
w_t=-\frac{w^{3/2}w_{xxx}}{w_x^3}
+3\frac{w^{3/2}w_{xx}^2}{w_x^4}
-\frac{3}{2}
\frac{w^{1/2} w_{xx}}{w_x^2}.
\end{gather}

\end{itemize}

} 

\noindent
We obtained one multipotentialisation for equation (\ref{Case-III-alpha-0-beta-1-eq}) which is a result of the potentialisation of equation (\ref{Case-III-Multi-x}) viz.
\begin{gather*}
v_t=\frac{v_x^2v_{xxx}}{v_{xx}^3}-\frac{2}{9}x,
\end{gather*}
which we now discuss in detail. 

\strut\hfill

\noindent
Equation (\ref{Case-III-Multi-x}) admits the following four integrating factors (no zero-order integrating factor exists):
\begin{subequations}
\begin{gather}
\label{L-III-Gen-4-1}
^{III}\Lambda^4_1[v]=
\frac{v_x^{2/3}v_{4x}}{v_{xx}^3}
-3\frac{v_x^{2/3}v_{xxx}^2}{v_{xx}^4}
+\frac{4}{3}\frac{v_{xxx}}{v_{xx}^2v_x^{1/3}}
+\frac{2}{9}\frac{1}{v_x^{4/3}}
\\[0.3cm]
\label{L-III-Gen-4-2}
^{III}\Lambda^4_2[v]=
\frac{v_{xx}}{v_x^{4/3}}
\\[0.3cm]
\label{L-III-Gen-4-3}
^{III}\Lambda^4_3[v]=
\frac{v_{xx}}{v_x^{5/3}}
\\[0.3cm]
\label{L-III-Gen-4-4}
^{III}\Lambda^4_4[v]=
v_{xx}
\end{gather}
\end{subequations}

\noindent
The corresponding conserved currents and fluxes for (\ref{Case-III-Multi-x}) are as follows:
\begin{subequations}
\begin{gather}
\label{Phi-t-III-lam-4-1}
^{III}\Phi^t_{4,1}[v]=\frac{v_x^{2/3}}{v_{xx}}
\\[0.3cm]
\label{Phi-x-III-lam-4-1}
^{III}\Phi^x_{4,1}[v]=\frac{v_x^{8/3}v_{4x}}{v_{xx}^5}
+\frac{2}{3}\frac{v_x^{5/3}v_{xxx}}{v_{xx}^4}
-2\frac{v_x^{8/3}v_{xxx}^2}{v_{xx}^6}
+\frac{2}{9}\frac{v_x^{2/3}}{v_{xx}^2}
\\[0.3cm]
\label{Phi-t-III-lam-4-2}
^{III}\Phi^t_{4,2}[v]=\frac{9}{4}v_x^{2/3}
\\[0.3cm]
\label{Phi-x-III-lam-4-2}
^{III}\Phi^x_{4,2}[v]=-\frac{3}{2}\frac{v_x^{5/3}v_{xxx}}{v_{xx}^3}
+\frac{1}{2}\frac{v_x^{2/3}}{v_{xx}}
\\[0.3cm]
\label{Phi-t-III-lam-4-3}
^{III}\Phi^t_{4,3}[v]=3v_x^{1/3}
\\[0.3cm]
\label{Phi-x-III-lam-4-3}
^{III}\Phi^x_{4,3}[v]=-\frac{v_x^{4/3}v_{xxx}}{v_{xx}^3}
+\frac{2}{3}\frac{v_x^{1/3}}{v_{xx}}
\\[0.3cm]
\label{Phi-t-III-lam-4-4}
^{III}\Phi^t_{4,4}[v]=\frac{1}{64}v_x^2
\\[0.3cm]
\label{Phi-x-III-lam-4-4}
^{III}\Phi^x_{4,4}[v]=-\frac{1}{32}\frac{v_x^3v_{xxx}}{v_{xx}^3}
-\frac{1}{32}\frac{v_x^2}{v_{xx}}
+\frac{5}{72}v.
\end{gather}
\end{subequations}

\noindent
This leads to 

\strut\vfill

\pagebreak

\noindent
{\bf Potentialisation III.2} {\it

\begin{itemize}

\item[a)]
Using the conserved current $^{III}\Phi^t_{4,1}$, given by (\ref{Phi-t-III-lam-4-1}), we find that equation 
(\ref{Case-III-Multi-x}) viz.
\begin{gather*}
\boxed{\vphantom{\frac{DA}{DB}}
v_t=\frac{v_x^2v_{xxx}}{v_{xx}^3}-\frac{2}{9}x
}\ ,
\end{gather*}
admits the zero-order potentialisation 
\begin{gather}
\label{EQ3}
\tilde v_{1,t}=\tilde v_{1,x}^3\tilde v_{1,xxx},
\end{gather}
where 
\begin{gather}
\tilde v_{1,x}=\frac{v_x^{2/3}}{v_{xx}},
\end{gather}
and the first-order potentialisation with
$\tilde w_{1,x}=D_x(^{III}\Phi^t_{4,1})$, i.e. $\tilde v_{1,x}=\tilde w_1$, namely
\begin{gather}
\label{EQ12}
\tilde w_{1,t}=\tilde w_1^3\tilde w_{1,xxx}
+3\tilde w_1^2\tilde w_{1,x}\tilde w_{1,xx}.
\end{gather}
Equation (\ref{EQ3}) also admits the zero-order potentialisation 
\begin{gather}
\label{EQ4}
\tilde w_{3,t}=
\frac{\tilde w_{3,xxx}}{\tilde w_{3,x}^3}-3
\frac{\tilde w_{3,xx}^2}{\tilde w_{3,x}^4},
\end{gather}
where 
\begin{gather}
\tilde w_{3,x}=\frac{1}{\tilde v_{1,x}}
\end{gather}
which corresponds to the integrating factor $\Lambda[\tilde v_1]=-2\tilde v_{1,x}^{-3}v_{1,xx}$ admitted by equation (\ref{EQ3}).

\item[b)]
Using the conserved current $^{III}\Phi^t_{4,2}$, given by (\ref{Phi-t-III-lam-4-2}), we find that equation 
(\ref{Case-III-Multi-x}) viz.
\begin{gather*}
\boxed{\vphantom{\frac{DA}{DB}}
v_t=\frac{v_x^2v_{xxx}}{v_{xx}^3}-\frac{2}{9}x
}\ ,
\end{gather*}
admits the zero-order potentialisation 
\begin{gather}
\label{EQ6}
\tilde v_{2,t}=\frac{\tilde v_{2,x}^{3/2}\tilde v_{2,xxx}}{\tilde v_{2,xx}^3}
\end{gather}
where 
\begin{gather}
\tilde v_{2,x}=\frac{9}{4}v_{x}^{2/3},
\end{gather}
and the first-order potentialisation with
$\tilde w_{2,x}=D_x(^{III}\Phi^t_{4,2})$, i.e. $\tilde v_{2,x}=\tilde w_2$, namely
\begin{gather}
\label{EQ7}
\tilde w_{2,t}=
\frac{\tilde w_2^{3/2}\tilde w_{2,xxx}}{\tilde w_{2,x}^3}
-3\frac{\tilde w_2^{3/2}\tilde w_{2,xx}^2}{\tilde w_{2,x}^4}
+\frac{3}{2}
\frac{\tilde w_2^{1/2}\tilde w_{2,xx}}{\tilde w_{2,x}^2}.
\end{gather}

\item[c)]
Using the conserved current $^{III}\Phi^t_{4,3}$, given by (\ref{Phi-t-III-lam-4-3}), we find that equation 
(\ref{Case-III-Multi-x}) viz.
\begin{gather*}
\boxed{\vphantom{\frac{DA}{DB}}
v_t=\frac{v_x^2v_{xxx}}{v_{xx}^3}-\frac{2}{9}x
}\ ,
\end{gather*}
admits the zero-order potentialisation 
\begin{gather}
\label{EQ8}
\tilde v_{3,t}=\frac{\tilde v_{3,xxx}}{\tilde v_{3,xx}^3},
\end{gather}
where 
\begin{gather}
\tilde v_{3,x}=3v_{x}^{1/3}
\end{gather}
and the first-order potentialisation with
$\tilde w_{3,x}=D_x(^{III}\Phi^t_{4,3})$, i.e. $\tilde v_{3,x}=\tilde w_3$, namely
\begin{gather*}
\tilde w_{3,t}=
\frac{\tilde w_{3,xxx}}{\tilde w_{3,x}^3}
-3\frac{\tilde w_{3,xx}^2}{\tilde w_{3,x}^4},
\end{gather*}
which is equivalent to (\ref{EQ4}).
Equation (\ref{Case-III-Multi-x}) also admits the second-order potentialisation with $W_{3,x}=D_x^2(^{III}\Phi^t_{4,3})$, i.e. $\tilde w_{3,x}= W_3$, namely
\begin{gather}
\label{EQ5}
W_{3,t}=
\frac{W_{3,xxx}}{W_3^3}
-9\frac{W_{3,x}W_{3,xx}}{W_{3}^4}
+12\frac{W_{3,x}^3}{W_3^5}.
\end{gather}

\item[d)]
Using the conserved current $^{III}\Phi^t_{4,4}$, given by (\ref{Phi-t-III-lam-4-4}), we find that equation 
(\ref{Case-III-Multi-x}) viz.
\begin{gather*}
\boxed{\vphantom{\frac{DA}{DB}}
v_t=\frac{v_x^2v_{xxx}}{v_{xx}^3}-\frac{2}{9}x
}\ ,
\end{gather*}
admits no zero-order potentialisation. Instead (\ref{Case-III-Multi-x}) admits the first-order potentialisation with
$\tilde w_{4,x}=D_x(^{III}\Phi^t_{4,4})$, i.e. 
\begin{gather}
\tilde w_4=
\frac{1}{64} v_x^2
\end{gather}
namely
\begin{gather}
\label{EQ11}
\tilde w_{4,t}=
-\frac{\tilde w_4^{5/2}\tilde w_{4,xxx}}{\tilde w_{4,x}^3}
+3\frac{\tilde w_4^{5/2}\tilde w_{4,xx}^2}{\tilde w_{4,x}^4}
- \frac{5}{2}
\frac{\tilde w_4^{3/2}\tilde w_{4,xx}}{\tilde w_{4,x}^2}
+\frac{5}{9}w_4^{1/2}.
\end{gather}
\end{itemize}

\noindent
A graphical description of the above is given in Diagram 4.

}


\strut\hfill

\begin{displaymath}
\xymatrix{
\qquad &\qquad \qquad \qquad\mbox{{\bf Diagram 4 }}& \qquad\qquad & \qquad \qquad\qquad\qquad\\
& 
\boxed{\vphantom{\frac{DA}{DB}}
\mbox{(\ref{EQ5}) $W_3$} 
}
& & \\
& \ar[u]^{{\small \mbox{1st-order}}}
\boxed{\vphantom{\frac{DA}{DB}}
\mbox{(\ref{EQ4}) $\tilde w_3$} 
}
& 
\framebox{$
\displaystyle{
u_t=\frac{1}{u_{xxx}^2}} $}
\ar[d]_{{\small \mbox{zero-order}}}
\\
\boxed{\vphantom{\frac{DA}{DB}}
\mbox{(\ref{EQ12}) $\tilde w_1$} 
}
& \ar[l]_{{\small \mbox{1st-order}}\ \ }
\ar[u]^{\ \ \qquad{\small \mbox{zero-order}}}
\framebox{$
\displaystyle{
\tilde v_{1,t}=\tilde v_{1,x}^3\tilde v_{1,xxx}} $}
&\ar[dl]_{\ \ {\small \mbox{zero-order\ \ \ \ }}}
\ar[l]_{{\small \mbox{zero-order}}}   
\framebox{$
\displaystyle{v_t=\frac{v_x^2v_{xxx}}{v_{xx}^3}-\frac{2}{9}x} $}
\ar[r]^{\quad\ {\small \mbox{1st-order}}}
\ar[dr]^{{\ \ \ \small \mbox{zero-order}}}
& 
\boxed{\vphantom{\frac{DA}{DB}}
\mbox{(\ref{EQ11}) $\tilde w_4$} 
}
\\
& 
\framebox{$
\displaystyle{
\tilde v_{2,t}=\frac{\tilde v_{2,x}^{3/2}\tilde v_{2,xxx}}{\tilde v_{2,xx}^3}} $}
\ar[d]^{{\small \mbox{1st-order}}}
&  & 
\framebox{$
\displaystyle{
\tilde v_{3,t}=\frac{\tilde v_{3,xxx}}{\tilde v_{3,xx}^3}} $}
\ar[d]_{{\small \mbox{1st-order}}}
\\\
& 
\boxed{\vphantom{\frac{DA}{DB}}
\mbox{(\ref{EQ7}) $\tilde w_2$} 
}
& & 
\boxed{\vphantom{\frac{DA}{DB}}
\mbox{(\ref{EQ4}) $\tilde w_3$} 
}
\ar[d]_{{\small \mbox{2nd-order}}}
\\
& & & 
\boxed{\vphantom{\frac{DA}{DB}}
\mbox{(\ref{EQ5}) $W_3$} 
}
}
\end{displaymath}

\strut\hfill

\noindent
{\bf Case IV:} We consider (\ref{Case-IV-eq}), viz.
\begin{gather*}
u_t=\frac{4u_x^5}{(2b\,u_x^2-2u_xu_{xxx}+3u_{xx}^2)^2},
\end{gather*}
where $b$ is an arbitrary constant. Equation (\ref{Case-IV-eq}) is Möbius-invariant \cite{E-E-April2019}
and can therefore conveniently be expressed in terms of the Schwarzian derivative $S$, namely
\begin{gather}
\label{Case-IV-S}
u_t=\frac{u_x}{(b-S)^2},
\end{gather}
where $S$ is given by (\ref{Schwarzian}). Equation (\ref{Case-IV-S}) admits the following fourth-order integrating factors for arbitrary constant $b$ (no zero-order or second-order integrating factors exist):
\begin{subequations}
\begin{gather}
\label{L-IV-a}
^{IV}\Lambda_1[u]=\frac{S_x}{u_x}\\[0.3cm]
\label{L-IV-b}
^{IV}\Lambda_2[u]=\left(
\frac{u_{xx}^2}{u_x^4}
+\frac{2b}{u_x^2}\right)S_x
+\frac{2u_{xx}}{u_x^3}S^2
-\frac{4bu_{xx}}{u_x^3}S
+\frac{2b^2 u_{xx}}{u_x^3}.
\end{gather}
\end{subequations}
For the case where $b=0$, equation
\begin{gather}
\label{Case-IV-S-b=0}
u_t=\frac{u_x}{S^2},
\end{gather}
admits, in addition to $^{IV}\Lambda_1[u]_{b=0}$ 
and $^{IV}\Lambda_2[u]_{b=0}$ given (\ref{L-IV-a}) and (\ref{L-IV-b}) respectively, also the integrating factor
\begin{gather}
\label{L-IV-c}
^{IV}\Lambda_3[u]=
\left(\frac{uu_{xx}^2 }{u_x^4} 
-\frac{2u_{xx}}{u_x^2}\right)S_x
+\left(\frac{2uu_{xx}}{u_x^3}
-\frac{2}{u_x}
\right)S^2.
\end{gather}
Corresponding to (\ref{L-IV-a}) and (\ref{L-IV-b}), the respective conserved current and flux for (\ref{Case-IV-S})
are as follows:
\begin{subequations}
\begin{gather}
\label{Case-IV-1-Phi-t}
^{IV}\Phi^t_1[u]=\frac{1}{4}\frac{u_{xx}^2}{u_x^2}
\\[0.3cm]
^{IV}\Phi^x_1[u]=-\frac{1}{4}\frac{u_{xx}^2}{(S-b)^2u_x^2}
+\frac{u_{xx}S_x}{(S-b)^3u_x}
+\frac{1}{2}\frac{2S-b}{(S-b)^2}
\\[0.3cm]
\label{Case-IV-2-Phi-t}
^{IV}\Phi^t_2[u]=\frac{b u_{xx}^2}{u_x^3}
+\frac{1}{12}
\frac{u_{xx}^4}{u_x^5}-\frac{b^2}{u_x}
\\[0.3cm]
^{IV}\Phi^x_2[u]=-\frac{1}{12}
\frac{u_{xx}^4}{u_x^5(S-b)^2}
+\frac{2}{3}
\frac{u_{xx}^3S_x}{u_x^4 (S-b)^3}
+\frac{u_{xx}^2(2S-3b)}{u_x^3(S-b)^2}
+\frac{4bu_{xx}S_x}{u_x^2(S-b)^3}\nn\\[0.3cm]
\qquad
+\frac{(2S-b)^2}{u_x(S-b)^2}.
\end{gather}
\end{subequations}
For the case $b=0$ corresponding to the integrating factor (\ref{L-IV-c}), we obtain an additional conserved current and flux for equation
(\ref{Case-IV-S-b=0}), namely
\begin{subequations}
\begin{gather}
\label{Case-IV-3-Phi-t}
^{IV}\Phi^t_3[u]=
\frac{1}{12}
\frac{uu_{xx}^4}{u_x^5}
-\frac{1}{3}\frac{u_{xx}^3}{u_x^3}
\\[0.3cm]
^{IV}\Phi^x_3[u]=
-\frac{1}{12}
\frac{uu_{xx}^4}{u_x^5}\frac{1}{S^2}
+\frac{1}{3}
\frac{u_{xx}^3}{u_x^4}\,\frac{2uS_x+u_xS}{S^3}
-\frac{2u_{xx}^2}{u_x^3}\,\frac{u_xS_x-uS^2}{S^3}
-\frac{4u_{xx}}{u_x}\frac{1}{S}\nn\\[0.3cm]
\qquad
+\frac{4u}{u_x}.
\end{gather}
\end{subequations}

\noindent
This leads to

\strut\hfill

\noindent
{\bf Potentialisation IV} {\it 
Using the conserved current $^{IV}\Phi^t_{1}$, given by (\ref{Case-IV-1-Phi-t}), we find that equation 
(\ref{Case-IV-S}) viz.
\begin{gather*}
\boxed{\vphantom{\frac{DA}{DB}}
u_t=\frac{u_x}{(b-S)^2}
}\ ,
\end{gather*}
admits, for arbitrary constant $b$, the zero-order potentialisation 
\begin{gather}
\tilde v_{t}=
\frac{1}{\left(-v_{xx}+bv_x^{1/2}+v_x^{3/2}\right)^3}
\left[
2v_x^{3/2}v_{xxx}
-\frac{3}{2}v_x(6v_x+b)v_{xx}\right.
\nn\\[0.3cm]
\qquad
\left.
+\frac{1}{2}(6v_x+b)(2v_x+b)v_x^{3/2}
\right],
\end{gather}
where 
\begin{gather}
v_{x}=\frac{1}{4}\frac{u_{xx}^2}{u_x^2}
\end{gather}
and the first-order potentialisation with
$w_{x}=D_x(^{IV}\Phi^t_{1})$, i.e. $v_{x}= w$, namely
\begin{gather}
w_{t}=
\frac{1}{\left(
-w_x+bw^{1/2}+2w^{3/2}\right)^4}
\left[
-2w^{3/2}w_xw_{xxx}
+2w^2(2w+b)w_{xxx}
+6w^{3/2}w_{xx}^2\right.\nn\\[0.3cm]
\qquad
-3w^{1/2}w_x^2w_{xx}
-3w(10w+b)w_xw_{xx}
+\frac{3}{2}(12w+b)w_x^3
+4w^{3/2}(6w-b)w_x^2\nn\\[0.3cm]
\qquad
-w^2(2w+b)(6w-b)w_x
\left.
\vphantom{w^{3/2}}
\right].
\end{gather}
No further potentialisations of any order are possible for equation (\ref{Case-IV-S}) or equation 
(\ref{Case-IV-S-b=0}) using the conserved currents $^{IV}\Phi^t_{2}$ and $^{IV}\Phi^t_{3}$ given by 
(\ref{Case-IV-2-Phi-t}) and (\ref{Case-IV-3-Phi-t}), respectively.
}

\renewcommand{\theequation}{\thesection.\arabic{equation}}
\setcounter{equation}{0}

\section{Concluding remarks}

In this article we are reporting all potentialisations of the class of fully-nonlinear symmetry-integrable equatons listed in Proposition 1 using their integrating factors up to order four, whereby the integrating factors do not depend explicitly on their independent variables. Several mappings to equations using the multi-potentialisations process are also given where possible, although we do not claim to have obtained here a complete list of all possible multi-potentialisations connected to the equations in Proposition 1.

Our results show that the class of equations in Proposition 1 have a rich structure and the systematic use of the zero and higher-order potentialisations, as introduced here, leads to interesting quasi-linear equations, some of which certainly deserve further attention.  

We should point out that in all our previous classifications of evolution equations of order three and order five where we have introduced potentialisations were based only on zero-order potentialisations. However, in the current work we have shown that there are many cases where an equation does not admit a zero-order potentialisation but that it can instead admit first and higher-order potentialisations. This means that some of those earlier classifications could possibly be extended by considering higher-order potentialisations. We will address this in the near future.

\label{lastpage}

\end{document}